\documentclass[galaxies,review,accept,moreauthors,pdftex,10pt,a4paper]{Definitions/mdpi} 

\def\lsim{~\rlap{$<$}{\lower 1.0ex\hbox{$\sim$}}}
\def\bsim{~\rlap{$>$}{\lower 1.0ex\hbox{$\sim$}}}

\usepackage{mathrsfs}

\firstpage{1} 
\pubvolume{xx}
\issuenum{1}
\articlenumber{1}
\pubyear{2019}
\copyrightyear{2019}
\externaleditor{Academic Editor: Roberta Humphreys}
\history{Received: date; Accepted: date; Published: date}

\Title{Massive stars in the Tarantula Nebula: A Rosetta Stone for Extragalactic Supergiant HII 
Regions}

\Author{Paul A Crowther*\orcidA{}}

\AuthorNames{Paul A Crowther}

\address{%
Department of Physics \& Astronomy, University of Sheffield, Sheffield, S3 7RH, United Kingdom}

\corres{Correspondence: paul.crowther@sheffield.ac.uk}

\abstract{A review of the properties of the Tarantula Nebula (30 Doradus) in the Large Magellanic Cloud is presented, primarily from the perspective of its massive star content. The proximity of the Tarantula and its accessibility to X-ray through radio observations permit it to serve as a Rosetta Stone amongst extragalactic supergiant HII regions since one can consider both its integrated characteristics and the individual properties of individual massive stars. Recent surveys of its high mass stellar content, notably the VLT FLAMES Tarantula Survey (VFTS), are reviewed, together with VLT/MUSE observations of the central ionizing region NGC~2070 and HST/STIS spectroscopy of the young dense cluster R136, provide a near complete Hertzsprung-Russell diagram of the region, and cumulative ionizing output. Several high mass binaries are highlighted, some of which have been identified from a recent X-ray survey. Brief comparisons with the stellar content of giant HII regions in the Milky Way (NGC~3372) and Small Magellanic Cloud (NGC~346) are also made, together with Green Pea galaxies and star forming knots in high$-z$ galaxies. Finally, the prospect of studying massive stars in metal poor galaxies is evaluated.}


\begin{document}


\section{Introduction}

The Tarantula Nebula (alias 30 Doradus) in the Large Magellanic Cloud (LMC) is the brightest 
supergiant HII region in the Local Group of galaxies, and serves as a local analogue to metal-poor 
starburst knots in high redshift galaxies \cite{Kennicutt98}. Its proximity (50 kpc), and high 
galactic latitude (and hence low extinction) have permitted a myriad of ground-based and space-based 
surveys across the electromagnetic spectrum, revealing an exceptional population of massive stars {\bf ($\geq 8 M_{\odot}$)}, including  the dense, young star cluster R136 that is home to some of the most 
massive stars known. The advent of modern highly multiplexed spectrographs coupled with large 
ground-based telescopes, has permitted multi-epoch optical spectroscopic surveys of the massive 
star population of the Tarantula Nebula for the first time. The VLT FLAMES Tarantula Survey, 
hereafter VFTS \cite{Evans11}, has provided the multiplicity, rotational velocities and initial 
mass function of massive stars. In distant starburst regions it is not possible to resolve 
individual stars, such that studies rely on techniques based on their integrated properties, which 
themselves involve assumptions of binarity, stellar rotation and mass function.

\begin{figure}[htbp]
\centering
\includegraphics[width=15 cm]{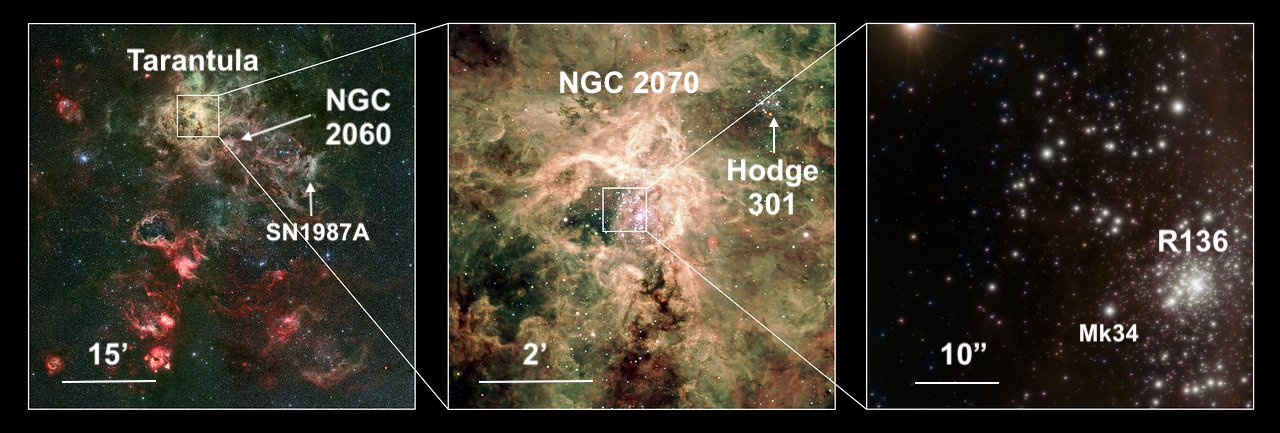}
\caption{(left) Optical image of the Tarantula Nebula from the MPG/ESO 2.2m WFI, with NGC~2060 and SN1987A indicated; (centre) Optical VLT/FORS2 image centred on NGC~2070, with Hodge~301 to the upper right; (right) an infrared VLT/MAD image of the central R136 region, with the massive colliding wind binary Mk~34 indicated. Credit: ESO/P. Crowther/C.J. Evans}\label{Tarantula}
\end{figure}


The Tarantula Nebula is not the sole example of a supergiant H\,{\sc ii} region within the Local 
Group. However, studies of the richest Milky Way star-forming regions are limited by dust 
extinction (e.g. Westerlund~1, NGC~3603). Counterparts in other Local Group galaxies suffer from a 
number of limitations, including a relatively modest stellar content (NGC~346 in the Small 
Magellanic Cloud, SMC) or much greater distance (NGC~604 in M\,33), such that the Tarantula -- 
whose metallicity is approximately half-solar \cite{Tsamis05}  -- serves as the only credible 
Rosetta Stone for rich extragalactic star-forming regions. This review will provide a brief 
overview of the structural properties of the Tarantula Nebula, but will largely focus on its 
massive star content, drawn from results from VFTS and other optical spectroscopic surveys, 
supplemented by the recent deep Chandra X-ray survey `Tarantula - Revealed by X-rays' (T-ReX). 
Finally, comparisons with local and high-redshift star-forming regions will be provided to put the 
properties of the Tarantula into a broader context. Indeed, the nebular properties of the central 
NGC~2070 region of the Tarantula are strikingly similar to Green Pea galaxies, exhibiting intense 
[O\,{\sc iii}] $\lambda$4959, 5007 emission, and its star-formation rate is comparable to intense 
star-forming clumps at high redshift.

\begin{table}[htbp]
\caption{Physical scales within the Tarantula Nebula, adapted from Walborn \cite{Walborn91}. Ionizing outputs, N(LyC) are obtained from the present work.}\label{scale} 
\centering
\begin{tabular}{l@{\hspace{-2mm}}r@{\hspace{2mm}}r@{\hspace{2mm}}r@{\hspace{2mm}}l@{\hspace{2mm}}l}
\toprule
\textbf{Region}	& \textbf{Angular}	& \textbf{Physical} & \textbf{N(LyC)} & \textbf{Content}  & \textbf{Reference}\\
                          & \textbf{radius ($''$)} & \textbf{radius (pc)} & \textbf{($10^{51}$ ph\,s$^{-1}$)} &  \\
\midrule
R136a               & 0.8       & 0.2  & 2 & R136a1 (WN5h), R136a2 (WN5h) & \cite{Crowther10} \\
R136                 &  4         & 1.0  & 4 & R136b (O4\,If/WN8), R136c (WN5h+) &  \cite{Crowther16} \\
NGC~2070        & 80       & 20.  & 9 & R140a (WC4+WN6+), Mk34 (WN5h+WN5h) & \cite{Castro18, Castro19} \\
Tarantula  &   600   & 150. & 12 & Hodge 301, PSR J0537-6910 (pulsar), N157B (SNR) & \cite{Doran13} \\
\bottomrule
\end{tabular}
\end{table}

\section{Tarantula Nebula}

The Tarantula Nebula is the most striking star-forming region in the LMC, whether viewed in the far 
ultraviolet (hot, luminous stars), H$\alpha$ (ionized gas) or mid-IR (warm dust), extending over 
several hundred parsec, owing to a massive stellar content producing an ionizing output which is a 
thousand-fold higher than the Orion Nebula {\bf \cite{Baldwin91}.}

NGC~2070 is the dominant ionized region within the Tarantula Nebula, powered by a large number of 
hot luminous stars from R136 at its heart plus many more within its vicinity. NGC~2060, located 6 
arcmin (90 pc) to the southwest, is host to a more modest population of OB stars plus an X-ray 
pulsar PSR J0537-6910 and its supernova remnant (SNR) N157B. Hodge~301 is located 3 arcmin (45 pc) to the 
north west of R136, but does not possess significant nebulosity since previous supernovae are 
likely to have cleared this region of gas, and its stellar population (B-type stars and red 
supergiants) does not possess a significant Lyman continuum output. NGC~2070 is often referred to 
as a cluster in the literature but it extends over tens of parsecs whereas genuine star clusters 
are an order of magnitude smaller, such that the only rich star clusters within the Tarantula are 
R136, with an age of 1--2 Myr \cite{Crowther16} and Hodge~301 with an age of 20--30 Myr 
\cite{Britavskiy19} with a few additional lower mass young, compact clusters (e.g. TLD1, SL 639). 
Table~\ref{scale} compares various regions within the Tarantula Nebula, adapted from a previous 
review by Walborn \cite{Walborn91}. Although the focus of the present review is on spectroscopic 
results for massive stars, Sabbi et al. \cite{Sabbi13} have undertaken a deep Hubble Space 
Telescope (HST) multi-colour photometric survey, known as the Hubble Tarantula Treasury Project 
(HTTP) which permits lower mass stars in the Tarantula Nebula to be studied.


A number of complementary studies of the star formation history of the Tarantula Nebula have been carried out, exploiting pre-main sequence 
low mass stars \cite{DeMarchi11, Sabbi16} and massive stars \cite{Schneider18b}. Significant star formation commenced $\sim$25 Myr ago, as 
witnessed by Hodge~301, and reached a peak several Myr ago, with the young massive star cluster R136 at its heart no more than 2 Myr old. It 
is apparent that star formation within the Tarantula Nebula has not been limited to specific parsec-scale star clusters, such as Hodge 301 or 
R136, but has been distributed across the entire region, akin to a super OB association. Wright et al. \cite{Wright16} have established from 
proper motion observations that star formation in the far smaller Milky Way Cygnus OB2 region did not originate in a star cluster, but 
involved individual sub-regions in virial equilibrium. Indeed, median ages of massive stars show little radial dependence on their projected 
distance from R136, with very massive stars ($\geq 100 M_{\odot}$) identified throughout the region \cite{Schneider18b}. Infrared and radio 
observations of the Tarantula reveal ongoing regions of massive star formation, to which the reader is referred to the review by Walborn 
\cite{Walborn02} and a more recent study of the brightest embedded sources based on Spitzer/IRAC imaging \cite{Walborn13}. Atacama Large 
Millimetre Array (ALMA) has obtained high resolution observations of parsec-scale clumps within the Tarantula \cite{Indebetouw13},  with the 
rate of star formation in the Tarantula anticipated to decline in the future.

\begin{table}[htbp]
\caption{Integrated nebular properties of nearby giant HII regions, adapted from Kennicutt \cite{Kennicutt84} for an assumed O7V Lyman continuum output of 10$^{49}$ ph\,s$^{-1}$.}\label{GHII} 
\centering
\begin{tabular}{lccc}
\toprule
\textbf{Region (galaxy)}	& \textbf{Diameter}	& \textbf{L(H$\alpha$)}\ & \textbf{N(O7V)} \\
                                           & \textbf{pc}              & \textbf{10$^{39}$ erg\,s$^{-1}$} & \\
\midrule
NGC~3372 (Milky Way)        & 200:                       & 0.8   & \phantom{00}45\\
NGC~346 (SMC)		& 220			& 0.8  &  \phantom{00}45\\
NGC~3603 (Milky Way)     & 100                        & 1.5 & \phantom{0}110 \\
NGC~604 (M33)                & 400                        & 4.5   & \phantom{0}320 \\
Tarantula (LMC)                &370                          & 15.\phantom{0}   & 1100 \\ 
\bottomrule
\end{tabular}
\end{table}

\section{Massive star content}

The integrated nebular properties of the Tarantula Nebula and other selected giant H\,{\sc ii} 
regions in the Local Group is presented in Table~\ref{GHII}. The ionizing output of the Tarantula 
Nebula corresponds to the equivalent of over a thousand O7V stars, each with 10$^{49}$ 
ph\,s$^{-1}$. In reality, the Tarantula hosts somewhat fewer O-type stars since the most extreme 
examples -- early O stars and luminous Wolf-Rayet (WR) stars -- each produce an order of magnitude 
more Lyman continuum photons. Nevertheless, this population represents an order of magnitude more O 
and WR stars than any Milky Way or SMC giant HII regions, and is not likely to be improved upon 
until extremely large telescopes are capable of resolving the massive stellar content of more 
extreme giant HII regions, such as NGC\,5461, 5462 and 5471 in M\,101 \cite{Kennicutt98}.

Historically, there have been several photometric and spectroscopic surveys of early-type stars in 
the Tarantula Nebula, each of which have employed contemporary (Galactic) temperature calibrations 
to produce Hertzsprung-Russell diagrams. Parker \cite{Parker93a, Parker93b} obtained the first 
extensive study of the entire region. Subsequently, Massey \& Hunter \cite{Massey98} 
obtained high-spatial 
resolution HST spectroscopy of early-type stars in the central, crowded R136 region, revealing an 
exceptional population of very early O stars, whilst Melnick and colleagues \cite{Bosch99, 
Selman99} obtained spectroscopy for a large number of early-type stars in the NGC~2070 region. The 
advent of efficient multi-object spectrographs on 8-10m telescopes, has permitted the most 
comprehensive optical 
spectroscopic survey of massive stars in the Tarantula Nebula to date through the VLT FLAMES 
Tarantula Survey (VFTS) \cite{Evans11}. Multi-object, multi-epoch spectroscopy of $\sim$800 massive 
stars across the entire region, for which detailed spectroscopic analyses have been undertaken for 
over 500 O and early B stars, such that temperature calibrations are no longer necessary. Although 
this represents the most extensive of early-type stars in a single star-forming region to date, 
this survey is incomplete owing the fibre-placement limitations, sampling $\sim$70\% of massive 
stars exterior to the dense R136 cluster from comparison with photometric surveys \cite{Doran13}.

\begin{table}[htbp]
\caption{Summary of stellar content of the Tarantula Nebula from recent spectroscopic surveys (excluding sources in common, although including individual components within SB2 binaries).}\label{massive_stars} 
\centering
\begin{tabular}{llcccccl}
\toprule
\textbf{Telescope/inst} & \textbf{Target} & \textbf{N(O-type)} & \textbf{N(B-type)} & \textbf{N(WR)}
& \textbf{N(Of/WN)} & \textbf{N(A+)} & \textbf{Reference} \\
\midrule
VLT/FLAMES  & 30 Dor            &    369             & 436     & 9  & 6     & 35    & \cite{Evans11} \\ 
HST/STIS      & R136                   &     57               & ..      & 3 & 2     & ..     & \cite{Crowther16} \\
VLT/MUSE     & NGC~2070       &    115              &  79     & .. & ..     & 1     & \cite{Castro18} \\ 
Other              & 30 Dor              &       29              &    8     &16 & ..    &  5  & \cite{Parker93b, Massey98, Bosch99, Breysacher99} \\
\midrule
Total               &  30 Dor              &       570            &  523   & 28  & 8   & 41  &   \\
\bottomrule
\end{tabular}
\end{table}

Two additional surveys have recently provided complete optical spectroscopic observations of all 
bright sources within the central crowded region of the Tarantula: (a) the central 4 arcsec (1 pc) 
of the R136 cluster exploiting the high spatial resolution of HST/STIS \cite{Crowther16}; (b) the 
central 2$\times$2 arcmin (30$\times$30 pc) region of NGC~2070 using the VLT/MUSE integral field 
spectrograph \cite{Castro18}. Although these lack the multi-epoch capabilities of VFTS, they are 
complementary since the richest stellar populations of the Tarantula are found within 
NGC~2070/R136, and provide extended wavelength coverage (yellow and red for MUSE, ultraviolet for 
STIS), albeit at reduced spectral resolution. A summary of these three contemporary surveys is 
provided in Table~\ref{massive_stars}, together with literature results. In total, approximately
1100 early-type massive stars have been spectroscopically observed. Our discussion of the massive star content of the Tarantula will largely draw upon results from VFTS, but include others where appropriate. It should be noted 
that analyses based on spectroscopic fibre-fed observations of early-type stars in regions of 
strong, highly variable nebulosity, is inherently problematic owing to the lack of local sky 
subtraction. Such issues do no affect long-slit STIS or integral field MUSE observations.

VFTS and other surveys have revealed that the Tarantula hosts extreme examples of stellar exotica, 
including the most massive stars known \cite{Crowther10}, a Luminous Blue Variable, R143 
\cite{Walborn17}, a very massive runaway \cite{Lennon18}, the fastest rotating stars 
\cite{Dufton11}, a massive overcontact binary \cite{Almeida15}, and a supernova remnant N157B 
\cite{Chen06}, with SN\,1987A located $\sim$300\,pc to the south west (Fig~\ref{Tarantula}).

\begin{figure}[htbp]
\centering
\includegraphics[width=7.5 cm]{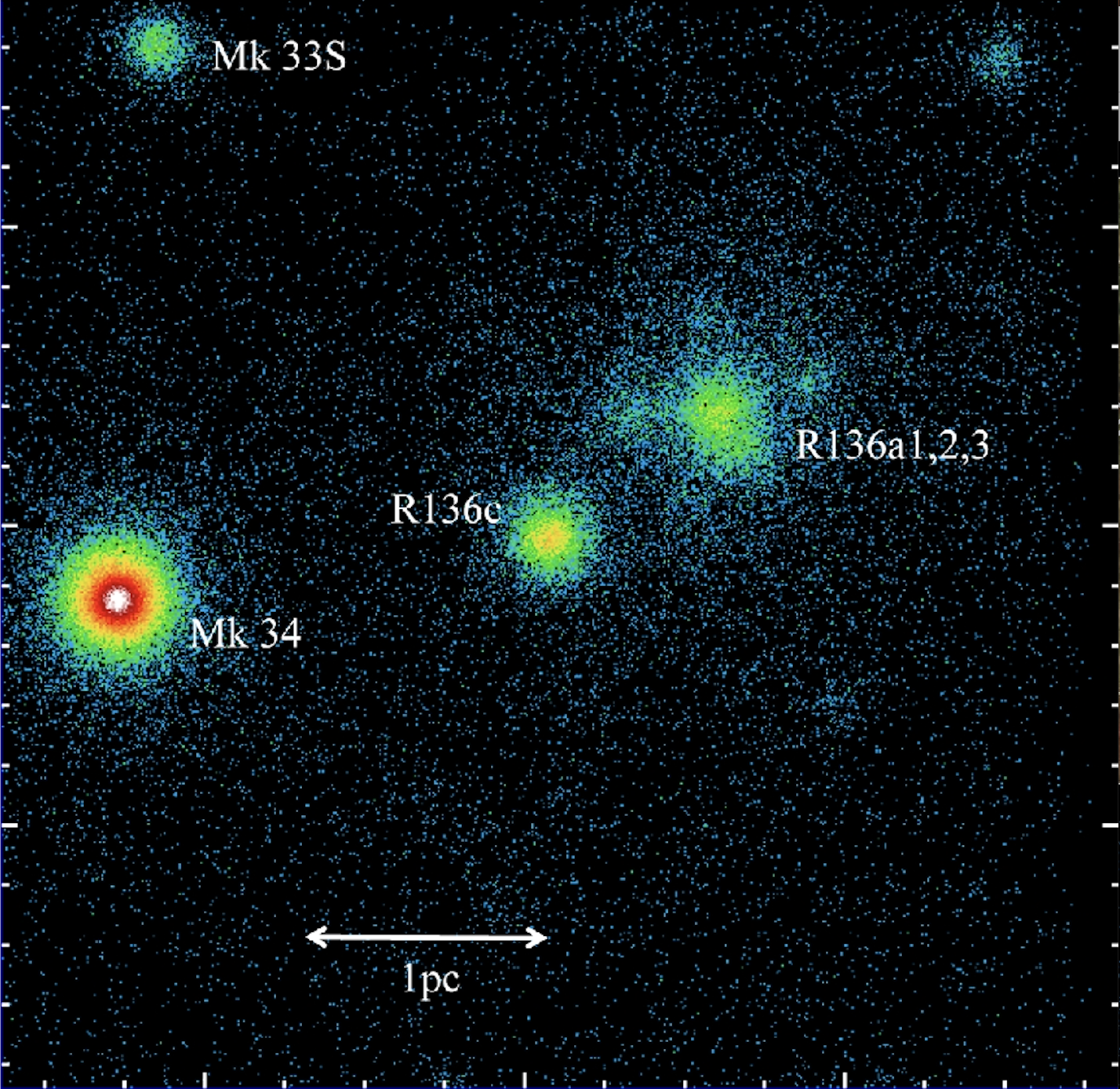}
\includegraphics[width=7.3 cm]{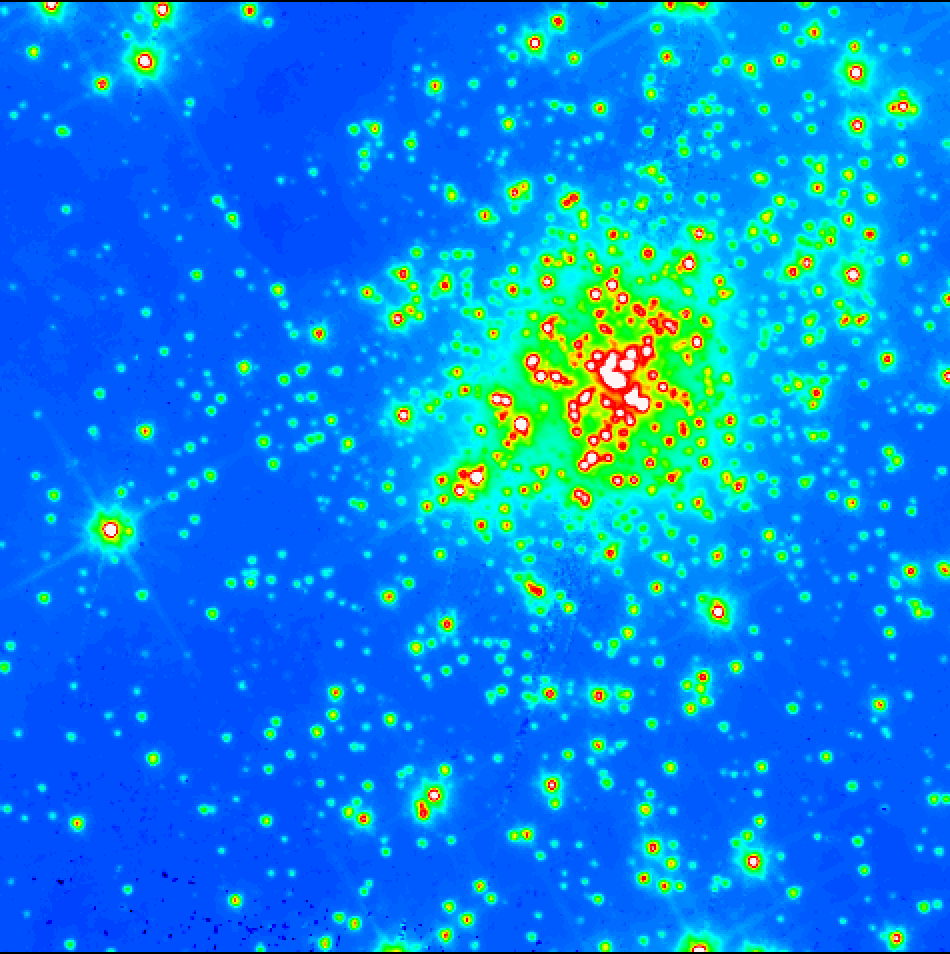}
\caption{(left) Chandra ACIS X-ray logarithmic intensity image of the core of NGC~2070 from T-ReX, centred on R136c, adapted from \cite{Pollock18}, showing the relative brightness of the colliding wind binary Melnick 34 (WNh5+WN5h) \cite{Tehrani19} to the R136a star cluster (hosting multiple WN5h stars) and R136c (WN5h+?); (right) HST WFC3/F555W logarithmic intensity image of the same 19$\times$19 arcsec region, highlighting the rich stellar population of R136a with respect to R136c and Melnick 34.}\label{TREX}
\end{figure}

R136a, the central cluster, merits special consideration since it was considered by some to be a 
supermassive star as recently as the early 1980s \cite{Savage83}. Speckle interferometry resolved 
R136a into multiple sources \cite{Weigelt85}, and it was subsequently established as a compact star 
cluster \cite{Hunter95}. Massey \& Hunter \cite{Massey98} established that dozens of the brightest 
sources within the central parsec were hot, early O stars. Spectroscopic studies of the brightest 
components R136a1, a2, a3, with nitrogen-sequence Wolf-Rayet spectral types, indicated masses of 
$\sim 100 M_{\odot}$ \cite{deKoter97, Massey98}. They established that these relatively weak-lined 
WN stars are luminous main-sequence stars close to their Eddington limits, rather than classical 
Wolf-Rayet stars. Subsequent analyses of the WN stars in R136 indicated significantly higher masses 
of 150--300 $M_{\odot}$ \cite{Crowther10} as a result of increased spectroscopic luminosities, 
owing to higher stellar temperatures and IR photometry less affected by dust extinction. Indirectly 
inferred masses of massive stars are notoriously imprecise, and if binarity were established for 
individual stars their inferred luminosities and masses would be reduced. To date, faint companions 
to members of R136a have been detected with extreme adaptive optics imaging \cite{Khorrami17}. 
Melnick 34 is spectroscopically similar to the WN5-stars in R136a, and has recently been shown to 
be a colliding wind binary system comprising two WN5 components, with a total mass exceeding 250 
$M_{\odot}$ \cite{Tehrani19}. Fig.~\ref{TREX} shows that Melnick 34 is an order of magnitude 
brighter than R136a in X-rays, indicating that there are no colliding wind binaries comparable to 
Melnick 34 within R136a \cite{Pollock18}.

Overall, the Tarantula hosts a remarkable number of $\sim$50 early-type stars with bolometric 
luminosities exceeding $10^{6} L_{\odot}$. For reference, the Milky Way's Carina Nebula (NGC~3372) 
hosts $\approx$5 massive stars with such extreme properties \cite{Smith06}. As such, the Tarantula 
Nebula represents our best opportunity to study the highest mass stars known, both individually and 
collectively. Schneider et al. \cite{Schneider18a} analysed VFTS spectroscopic results to establish 
an excess of massive stars with respect to a standard Salpeter Initial Mass Function (|MF), 
indicating 1/3 more stars with $\geq 30 M_{\odot}$ in 30 Doradus compared to expectations from a 
standard IMF. Finally, although we focus primarily on high mass early-type stars in the Tarantula 
Nebula, it also hosts red supergiants (RSG). Since RSG are the evolved descendants of moderately 
massive stars, and star formation in the Tarantula has peaked relatively recently, of order 
$\sim$10 RSG are known, most of which are associated with mature star clusters Hodge 301 and SL 639 
\cite{Britavskiy19}.


\section{Physical properties}

The determination of physical properties ($T_{\rm eff}, \log g$) of individual early-type stars 
ideally requires high S/N ($\geq$50) intermediate resolution spectroscopy of suitable diagnostics, 
usually He\,{\sc i-ii} lines (Si\,{\sc ii-iv}) for temperatures of O-type (B-type) stars, plus 
Balmer lines for surface gravities, plus grids of model atmospheres obtained with modern codes, 
such as FASTWIND \cite{Puls05} for O stars and blue supergiants, CMFGEN \cite{Hillier98} or PoWR 
\cite{Grafener02} for emission-line stars, or TLUSTY \cite{Lanz07} for low luminosity B stars. 
Analysis of late-type supergiants requires model atmosphere codes in which molecular opacities have 
been incorporated, such as MARCS \cite{Gustafsson08}. Hot O stars require alternative temperature 
diagnostics to helium, with nitrogen commonly used since the blue visual spectrum of O stars 
includes lines of N\,{\sc iii-v}. Wolf-Rayet stars are especially problematic since photospheres 
are masked by the dense wind, such that gravities cannot be directly measured and temperatures 
usually refer to deep layers, with an optical depth of $\tau \sim 10-20$, rather than the effective 
temperature at $\tau = 2/3$. If stellar distances are uncertain, comparisons with evolutionary 
models can be made using the so-called spectroscopic Hertzsprung-Russell (sHR) diagram 
\cite{Langer14}, involving temperature and ${\mathscr L }= T_{\rm eff}^{4} / g$, the inverse of the 
flux-weighted gravity, where $g$ is the surface gravity.

The determination of stellar luminosities requires comparisons between synthetic spectral energy 
distributions and photometry, taking account of interstellar extinctions and distance moduli, 18.5 
mag in the case of the Large Magellanic Cloud. Visual extinctions of early-type stars in the 
Tarantula Nebula are usually modest, although near-IR photometric comparisons usually lead to more 
robust luminosities since typical dust extinctions are 0.1--0.2 mag in the K-band, versus 1--2 mag 
in the V-band, and the lack of sensitivity of K-band extinctions to any variations in the overall 
extinction law. Luminosities of RSG can also be reliably estimated by integrating observed spectral 
energy distributions from visual to mid-IR wavelengths \cite{Gordon16, Davies18}. Historically, 
stellar estimates of masses and ages from evolutionary models involved by-eye comparisons between 
their position on a conventional Hertszprung-Russell (HR) diagram and theoretical isochrones. 
However, additional physical information is often available, such as helium abundance or projected 
rotational velocities. Tools now exist which additionally take such information into account for 
the calculation of stellar ages and initial masses such as BONNSAI \cite{Schneider14}. Significant 
discrepancies exist between current mass estimates from spectroscopic ($\log g$) and evolutionary 
approaches for a subset of VFTS O dwarfs \cite{SabinSanjulian17}.


\begin{figure}[htbp]
\centering
\includegraphics[width=14 cm]{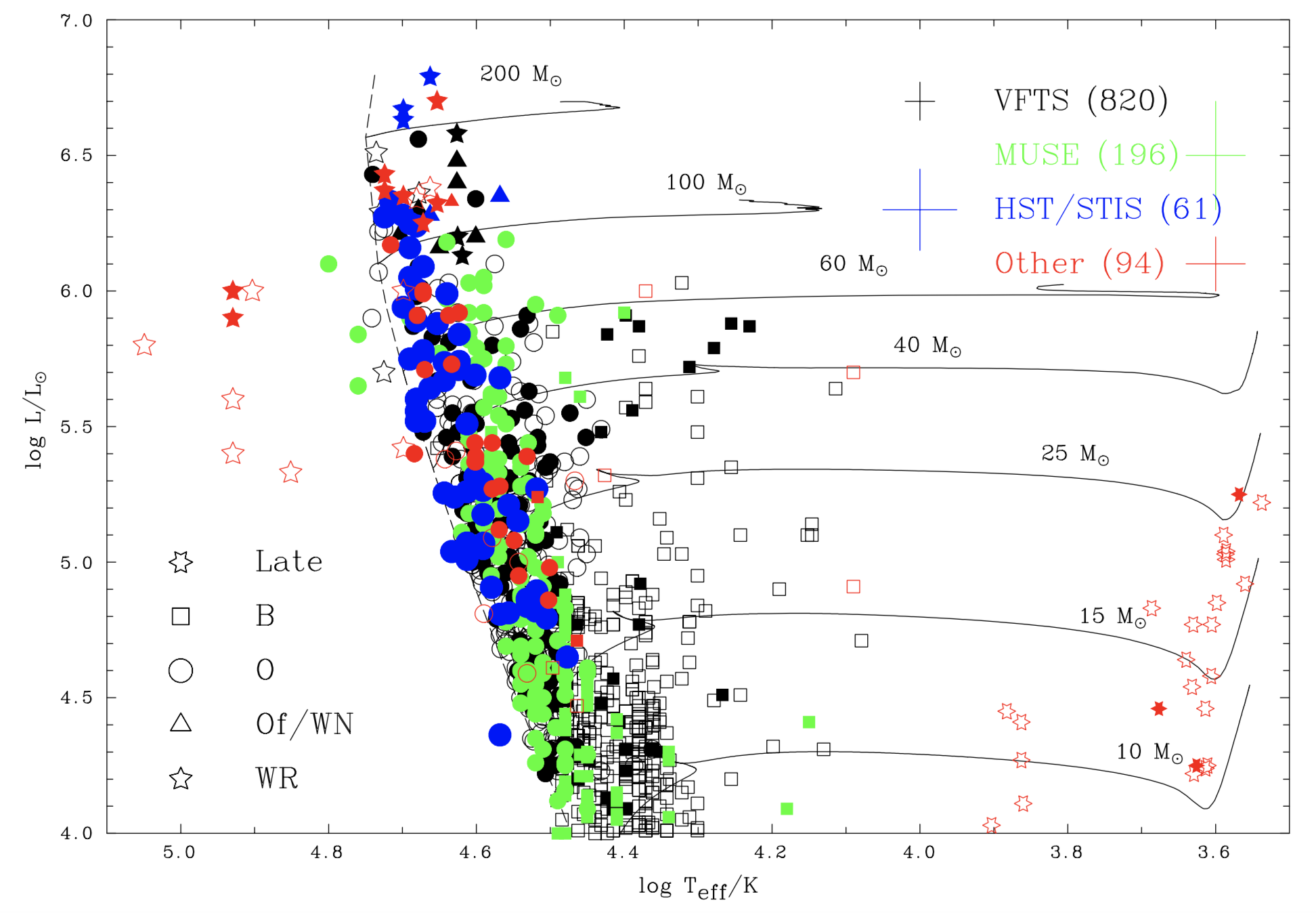}
\caption{Hertzsprung-Russell diagram of the Tarantula Nebula, based on results from VFTS 
\cite{Bestenlehner14, McEvoy15, SabinSanjulian17, RamirezAgudelo17, Dufton18}, MUSE 
\cite{Castro19}, HST/STIS \cite{Bestenlehner19b} and other literature results, with typical 
uncertainties from each survey indicated. Filled symbols are within NGC~2070, open symbols 
elsewhere in the Tarantula. Non-rotating tracks for 10, 15, 25, 40, 60, 100 and 200 $M_{\odot}$ LMC 
metallicity stars have been included from \cite{Brott11, Kohler15} which terminate at the onset of 
He-burning.}\label{HRD}
\end{figure}

Figure~\ref{HRD} presents the HR diagram of the Tarantula Nebula, comprising single star results 
from VFTS \cite{Bestenlehner14, McEvoy15, SabinSanjulian17, RamirezAgudelo17, Dufton18}, VLT/MUSE 
\cite{Castro19}, HST/STIS \cite{Bestenlehner19b} and literature results for other stars within 160 
parsec of R136, including Wolf-Rayet stars \cite{Hainich14}. Results for binary systems have 
been incorporated, primarily involving VFTS B-type binaries \cite{Garland17}, Tarantula 
Massive Binary Monitoring (TMBM) O-type binaries \cite{Mahy19} and recent literature for 
WR stars \cite{Shenar19, Tehrani19}. Evolutionary tracks for non-rotating, LMC metallicity massive 
stars up to the onset of He-burning have been included for reference \cite{Brott11, Kohler15}. Over 
1170 massive stars have been included, revealing a well populated main sequence population up to 
$\sim 200 M_{\odot}$, plus classical Wolf-Rayet stars to the left of the main sequence, and evolved 
blue supergiants up to $\log (L/L_{\odot}) \sim 6$, and cool supergiants, up to $\log (L/L_{\odot})
\sim$5.3 \cite{Davies18}. The addition of all luminous early-type stars from R136 and NGC~2070 
fills in the extreme upper main sequence which is somewhat under populated from VFTS alone 
\cite{Schneider18a}.  The overwhelming majority of the older massive stellar population -- i.e. 
evolved stars with masses below $30 M_{\odot}$ -- are spatially exterior to NGC~2070 (open 
symbols), although NGC~2070 is host to one luminous M supergiant, Melnick~9. Conversely, beyond 
NGC~2070, the main-sequence population at the highest stellar masses is relatively underpopulated, 
albeit with several WN5h stars (R146, R147) and early O stars (VFTS 16, BI 253) located 95$\pm$25 
parsec from R136.


\section{Binaries, rotation and runaways}

\begin{figure}[htbp]
\centering
\includegraphics[width=6.75 cm]{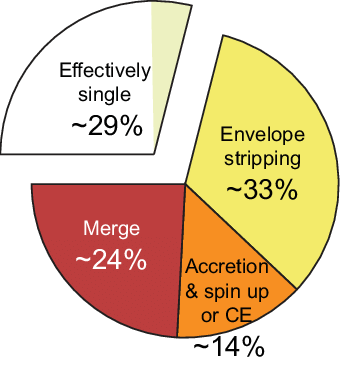}
\includegraphics[width=6 cm]{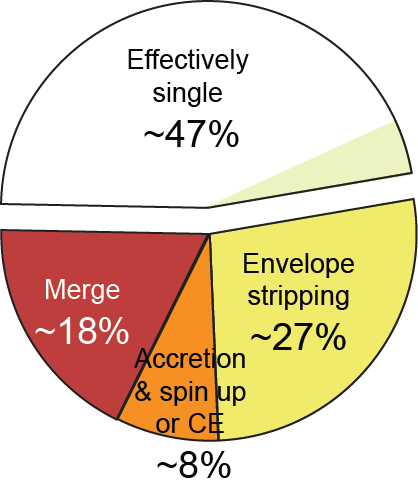}
\caption{Pie charts, courtesy Hugues Sana and Selma de Mink, illustrating the fraction of O stars undergoing single stellar evolution versus mergers, primaries being stripped of their envelopes, and secondaries being spun up, for Milky Way young clusters (left) and VFTS O stars (right), adapted from \cite{Sana12, Sana13}.}\label{piechart}
\end{figure}

Until recently, the significance of close binary evolution for massive stars was not fully 
recognised, in spite of a few binary ``champions" \cite{Vanbeveren98}. The high frequency of close 
binaries amongst O stars in young Galactic clusters obtained from radial velocity monitoring 
established that only a minority of massive stars follow single stellar evolution \cite{Sana12}. In 
contrast with the majority of previous spectroscopic surveys of early-type stars, VFTS comprised 
multiple epochs, such that \cite{Sana13} were able to establish that 53\% of O stars in the 
Tarantula Nebula inhabit a binary system with a period below 1,500 days, such that binary 
interaction will occur. 18\% of O-type binaries, those with very short periods are anticipated to 
merge with a companion, 27\% will be stripped of their envelopes (primaries, mass donors) and 8\% 
are predicted to be spun up (secondaries, mass gainers), as summarised in Figure~\ref{piechart} 
together with counterparts in Milky Way clusters. Broadly similar results have been obtained for 
VFTS B-type stars \cite{Dunstall15}.

The inferred rate of envelope stripping and spin-up in the Tarantula is rather lower than 
\cite{Sana12} obtained for O stars in Milky Way clusters (Fig.~\ref{piechart}), but it is probable 
that the true incidence is rather higher since a subset of the current O star sample is likely to 
have already undergone binary evolution. Although VFTS has established the binary frequency amongst 
massive stars in the Tarantula, binary orbits require follow-up surveys, notably the TMBM survey \cite{Almeida17, Mahy19}.

Consequences of close binary evolution include mass gaining secondaries being spun-up, and a subset 
of secondaries possessing high space velocities as a result of the disruption of the binary 
following the core-collapse supernova (ccSN) of the original primary. Other ``observables" are more 
challenging, including the identification of stripped primaries in close binaries which will 
usually be masked at visual wavelengths by mass gaining secondaries \cite{Gotberg18} which should 
be more common in older star clusters such as Hodge~301.

\begin{figure}[htbp]
\centering
\includegraphics[width=10 cm]{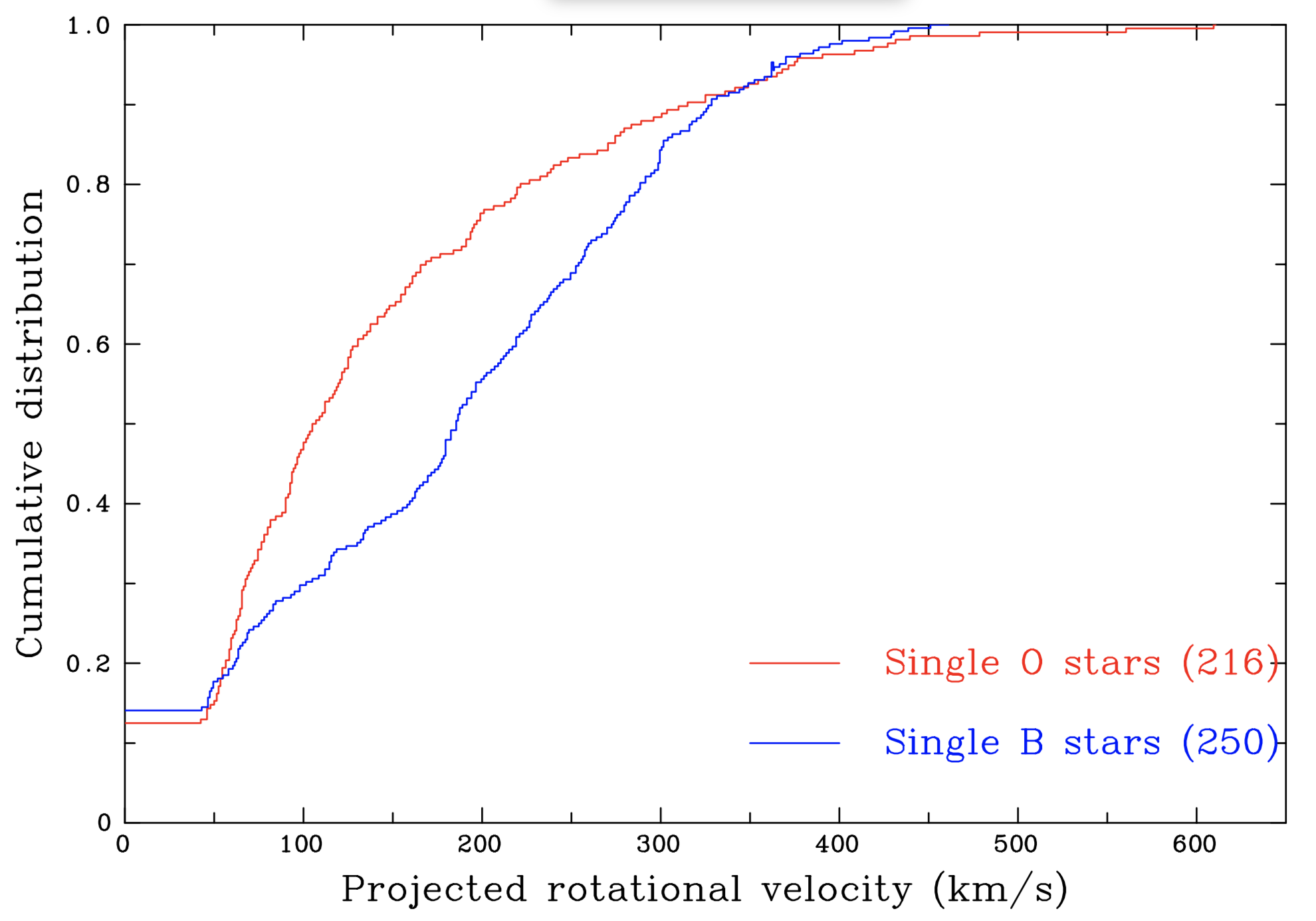}
\caption{Cumulative distribution of rotational rates for single VFTS O (red) and B stars (blue) \cite{RamirezAgudelo13, Dufton13}}\label{vsini}
\end{figure}

Reliable measurements of projected rotation rates, $v_{e} \sin i$, are often problematic because 
strong hydrogen and helium lines are predominantly affected by pressure broadening, with an 
additional contribution from``macroturbulence''. A Fourier Transform approach applied to metallic 
lines offers the most robust results, albeit requiring high resolution, high S/N spectroscopy of 
suitable diagnostics \cite{SimonDiaz14}. The lack of a spectral features originating in the 
hydrostatic layers of Wolf-Rayet stars prevents a direct determination of their rotational 
velocities.  Fig.~\ref{vsini} compares projected rotational velocities for a large sample of VFTS O and B-type stars. Typical rotational velocities of single O stars are modest, with $v_{e} \sin 
i \sim$ 100 km\,s$^{-1}$, albeit with 10\% exceeding 300 km\,s$^{-1}$ \cite{RamirezAgudelo13}, of 
which some examples are rotating close to their critical rates \cite{Dufton11}. This high velocity 
tail is suspected of being spun-up mass gainers in former close binaries. The lack of fast rotators 
amongst O stars in VFTS binary systems supports this interpretation \cite{RamirezAgudelo15}. 
Fig.~\ref{vsini} illustrates that rotational velocities of VFTS B-type stars are higher, with $v_{e} \sin i \sim$ 200 km\,s$^{-1}$ on average, and 20\% exceeding 300 km\,s$^{-1}$ \cite{Dufton13}.

Of particular interest is the rotational velocity distribution of massive stars within the young 
R136 star cluster, whose severe crowding prevented inclusion in VFTS. Bestenlehner et al. 
\cite{Bestenlehner19b} utilised HST/STIS spectroscopy to reveal 150 km\,s$^{-1}$ on average for a 
sample of 55 massive stars within the central parsec of R136, with no examples exceeding 250 
km\,s$^{-1}$, although the low S/N of these datasets prevented distinguishing between rotational 
and macroturbulence, adding to the interpretation that rapid rotators originate from close binary 
evolution. Wolff et al. \cite{Wolff08} obtained somewhat higher rotational velocities for OB stars 
in the periphery of R136, where contamination from the field population is significant.

The Tarantula hosts a number of candidate early-type runaway stars from their measured (radial 
and/or tangential) velocities with respect to the average for their environment, although radial 
velocity outliers may be unresolved binaries. Runaways can originate either from disrupted 
secondaries following the core collapse of primaries in close binaries, or following the dynamical 
ejection of stars from young star-forming regions. Platais et al. \cite{Platais18} have 
investigated high proper motion stars in the Tarantula from HST imaging obtained 3 years apart, 
revealing a number of potential stars ejected from R136, while Lennon et al. \cite{Lennon18} have 
exploited Gaia DR2 proper motions to conclude that VFTS 16 (O2\,III) \cite{Evans10} was likely 
ejected from R136 during its formation 1--2 Myr ago. Renzo et al. \cite{Renzo19} discuss the origin 
of the candidate `walkaway' very massive star VFTS 682 (WN5h).

The Tarantula hosts several notable massive binary systems, whose physical and orbital properties 
have been obtained from spectroscopic monitoring, with searches for massive binaries also greatly 
benefitting from the recent Chandra T-ReX survey (PI Leisa Townsley) which monitored the Tarantula 
in X-rays for almost 2 years with a total integration time of 2 Ms. Single hot, luminous stars tend 
to produce (thermal) X-rays due to shocks in the winds, but these are generally soft X-ray emitters 
with $L_{X}/L_{\rm Bol} \sim 10^{-7}$ \cite{Pallavicini81}. Massive stars in binary systems may
lead to excess X-ray emission arising from wind-wind collisions, usually relatively hard, providing 
the separations are not too small (low wind velocities) or too large (low wind densities) 
\cite{Stevens92}. A close binary comprising an early-type star and compact remnant (neutron star or 
black hole) will be extremely X-ray bright if the accretion disk of the remnant is being fed by the 
wind of the massive star or via Roche Lobe overflow.

A number of eclipsing binaries in the proximity of R136 have been identified \cite{Massey02}, 
including \# 38 from \cite{Hunter95} comprising an O3\,V+O6\,V in a circular 3.4 day orbit, with 
component masses 57 and 23 $M_{\odot}$. This represented the first robust stellar mass 
determination for an O3 star in the LMC. VFTS revealed a large number of binaries within the 
Tarantula, many of which have been followed-up with TMBM. Most notably R139 has been established as 
an eccentric system comprising a pair of mid O supergiants in a 154 day orbit \cite{Taylor11} with 
lower limits of $\sim 66+78 M_{\odot}$ for individual component, recently revised downward to 
$54+69 M_{\odot}$ \cite{Mahy19}. R139 is amongst the brightest X-ray sources in the Tarantula in 
T-ReX with $L_{\rm X, corr} \sim 5 \times 10^{33}$ erg\,s$^{-1}$ and an enhanced $L_{\rm X, 
corr}/L_{\rm Bol} \sim 9 \times 10^{-7}$ based on TMBM bolometric luminosities \cite{Mahy19}.

The most remarkable X-ray source in the Tarantula is VFTS 399 with $L_{\rm X, corr} \sim 5 \times 
10^{34}$ erg\,s$^{-1}$ despite being associated with a low luminosity O9 giant, implying $L_{\rm 
X,corr}/L_{\rm Bol} \sim 2 \times 10^{-4}$. Clark et al. \cite{Clark15} conclude that VFTS 399 is a 
high-mass X-ray binary hosting a neutron star remnant, with the O giant known to be a rapid rotator 
($v_{e} \sin i = 324$ km\,s$^{-1}$, according to \cite{RamirezAgudelo17}), as one would expect for 
a mass gainer in a close binary system.

Two point sources in the Tarantula are even brighter in X-rays than VFTS 399. Of these, R140a is a 
compact group of stars including two WR stars, so likely hosts one or more colliding wind binaries, 
while X-ray variability for Melnick~34 reveals a 155 day period, peaking at $L_{\rm X, corr} = 3.2 
\times 10^{35}$ erg\,s$^{-1}$, exceeding $\eta$ Carina at X-ray maximum \cite{Pollock18}. Melnick 
34 has been confirmed to be a colliding wind binary system in a 155 day eccentric orbit, with 
minimum masses of 60--65 $M_{\odot}$ for each of the WN5h components \cite{Tehrani19}. Individual 
masses of 130--140 $M_{\odot}$ are favoured from spectroscopic analysis, such that Melnick 34 is 
likely to be the most massive binary known to date, with $L_{\rm X, corr}/L_{\rm bol} = 1.7 \times 
10^{-5}$ at X-ray maximum. This arises from the collision of dense, fast moving winds at a minimum 
separation of $\sim$1.2 AU or 13 stellar radii for an assumed orbital inclination of $i = 
50^{\circ}$. Almeida et al. \cite{Almeida15} identify the overcontact binary VFTS 352 as a 
prototype of systems which, at low metallicity, are plausible black hole-black hole merger 
progenitors.

\begin{figure}[htbp]
\centering
\includegraphics[width=14 cm]{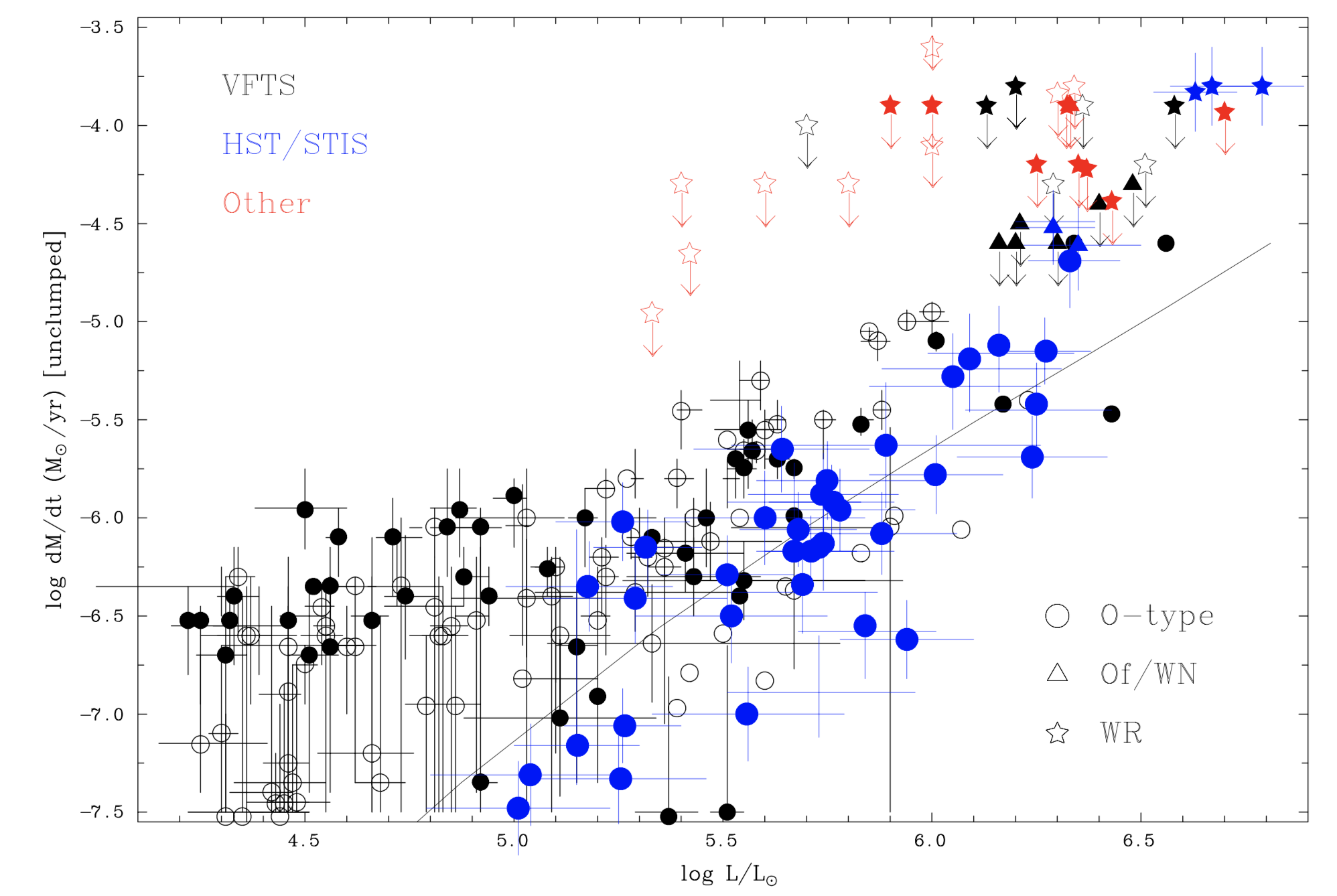}
\caption{Unclumped mass-loss rates of O-type, Of/WN and Wolf-Rayet stars in the Tarantula Nebula (based on results from VFTS \cite{Bestenlehner14, 
SabinSanjulian17, RamirezAgudelo17}, HST/STIS \cite{Bestenlehner19b} and other surveys \cite{Hainich14, Shenar19, Tehrani19} for WR stars). 
Filled symbols are within NGC~2070, open symbols elsewhere in the Tarantula. Theoretical mass-loss rates for zero age main sequence massive stars at the LMC metallicity \cite{Vink01} are included (solid line), based on LMC metallicity evolutionary models \cite{Brott11, Kohler15} }\label{Mdot}
\end{figure}

\section{Wind properties}

The underlying theory responsible for outflows by hot luminous stars has been known for several 
decades \cite{Castor75}, with a metallicity dependence primarily arising from the variation in 
iron-peak elemental abundances \cite{Vink01, Mokiem07}. Individual wind properties of O stars or 
blue supergiants usually rely on spectroscopic fits to H$\alpha$, as parameterised by the wind 
strength parameter, $Q$ \cite{Puls05}, from which the mass-loss rate requires knowledge of the 
physical radius (from $T_{\rm eff}, \log L$) and measured or adopted wind velocity. Wind velocities 
of OB stars cannot be measured from optical spectroscopy, so usually spectral type calibrations are 
adopted based on measured velocities from UV P Cygni profiles of C\,{\sc iv}, N\,{\sc v} or 
Si\,{\sc iv} \cite{Prinja90}. Until recently, high S/N, high quality UV spectroscopy of OB stars in 
the Large Magellanic Cloud has been in short supply, but the situation has improved via the HST 
Large Program METAL (GO 14675, PI Julia Roman-Duval) \cite{RomanDuval19} and upcoming ULLYSES 
initiative\footnote{http://www.stsci.edu/stsci-research/research-topics-and-programs/ullyses}.

Specifically for the Tarantula, low resolution UV spectroscopy of the R136 star cluster has added a 
significant number of wind measurements for early O stars \cite{Crowther16}. Wind velocities exceed 
3000 km\,s$^{-1}$ at the earliest subtypes (O2--3), reducing to $\sim$1500 km\,s$^{-1}$ for late 
O-types. Mass-loss rates of emission line stars typically rely on an alternative wind scaling 
relation, namely the transformed radius, $R_{t}$ \cite{Schmutz89}, which also requires knowledge of 
wind velocities, although these can be estimated from optical spectroscopy for Wolf-Rayet stars. An 
added complication arises because radiatively-driven winds are known to be  inherently unstable, 
leading to clumped winds. Wolf-Rayet winds have been known to be clumped for 30 years 
\cite{StLouis88, Hillier91}, but the degree of clumping for O stars via the H$\alpha$ diagnostic 
remains unclear, with conflicting results obtained from UV resonance lines \cite{Fullerton06} 
unless the wind comprises a mixture of optically thin and thick clumps within a much lower density 
inter-clump medium \cite{Sundqvist18}.

Fig~\ref{Mdot} presents unclumped mass-loss rates of O, Of/WN and Wolf-Rayet stars in the Tarantula 
Nebula, obtained from VFTS \cite{Bestenlehner14, SabinSanjulian17, RamirezAgudelo17}, HST/STIS 
\cite{Bestenlehner19b} and other literature results. Uncertainties have been included wherever 
possible. Since the primary wind diagnostic in the majority of instances presented here is 
H$\alpha$, it is apparent that uncertainties are large for those stars with weak stellar winds. In 
addition, mass-loss rates for Of/WN and Wolf-Rayet stars are anticipated to be reduced 
significantly owing to wind clumping, as indicated with downward arrows. If volume filling factors 
are $\sim$10\%, mass-loss rates will be reduced by a factor of $\sqrt{10}$.

Theoretical mass-loss rates \cite{Vink01} for zero-age main sequence stars at LMC composition 
\cite{Brott11, Kohler15} are included in Fig~\ref{Mdot}.  At face value it would appear that the 
theoretical mass-loss rates of LMC O stars are supported by theory. However, the following should 
be borne in mind. It is not clear how significantly wind clumping affects the inferred mass-loss 
rates of normal O stars, although H$\alpha$ results for supergiants are likely to be sensitive to 
wind clumping. In addition, the vast majority of mass-loss rates of VFTS O stars shown here have 
been inferred by adopting wind velocities from an assumed scaling relation involving escape 
velocities \cite{Lamers95, KudritzkiPuls00}, which are themselves dependent upon spectroscopic 
gravities. Exceptions are HST/STIS results for early-type stars in R136 which are based on measured 
UV wind velocities, and span dwarfs, giants, supergiants and main sequence WN stars. In order to 
verify predictions for lower luminosity ($\log L/L_{\odot} < 5.5$)  O stars, more sensitive 
diagnostics would need to be employed, such as UV P Cygni lines, providing complications such as 
porosity are accounted for \cite{Sundqvist18}.

It is clear from Fig.~\ref{Mdot} that rates for the highest luminosity main-sequence Of/WN and WN 
stars significantly exceed theoretical predictions. This discrepancy is partially addressed through 
wind clumping, but very massive stars close to their Eddington limits are observed to exhibit 
enhanced mass-loss rates which are not taken into account in standard theoretical predictions 
\cite{Grafener11, Bestenlehner19a}. Unsurprisingly, classical Wolf-Rayet stars with $\log (L/L_{\odot}) = 5.5-6$
possess the strongest winds amongst early-type stars in 30 Doradus, with clumping-corrected wind 
densities an order of magnitude higher than O stars with similar luminosities. It is well known 
that the wind momenta of WR stars, $\dot{M} v_{\infty}$, exceeds the momentum provided by their 
radiation field, $L/c$, owing to multiple photon absorption and re-emission within their optically 
thick winds, permitting $\dot{M} v_{\infty}/(L/c) > 1$ \cite{Puls08}. 

\section{Fate of massive stars in the Tarantula}

The conventional picture of massive star evolution close to solar metallicity is that those with 
initial masses of 8 -- 25 $M_{\odot}$ will end their lives as red supergiants (RSG), undergo a 
H-rich core collapse supernova, leaving behind a neutron star remnant, while higher mass 
counterparts will either circumvent the RSG phase or subsequently proceed to a Wolf-Rayet stage 
prior to undergoing core collapse, leading to a H-deficient supernova (neutron star remnant) or 
faint/failed supernova (black hole remnant) \cite{Langer12}. If one considers the global WR vs RSG 
population in the LMC, the lower boundary to the luminosity of WR stars is $\log (L/L_{\odot})$ = 
5.3, while the upper luminosity of RSG is $\log (L/L_{\odot})$ = 5.5 \cite{Davies18}, supporting a 
transition from RSG to WR for higher mass progenitors at $\sim$25 $M_{\odot}$.

Close binary evolution severely complicates this scenario, since primaries below 25 $M_{\odot}$ can 
be stripped of their hydrogen envelope, leading to a type IIb or Ib/c  instead of a H-rich 
supernova, while secondaries will be rejuvenated, spun-up, with the potential for a core-collapse 
supernova for secondaries whose initial masses fall below 8$M_{\odot}$. To date, there are no 
unambiguous cases of pre-supernova close binaries in the Tarantula hosting stripped (Wolf-Rayet or 
helium) stars, although it has been suggested that the WN3 binary BAT99-49 elsewhere in the LMC is 
the product of close binary evolution \cite{Shenar19}. Rapid rotation of the bright O giant 
component of the high mass X-ray binary VFTS 399 is consistent with this evolutionary scenario. The 
absence of low luminosity Wolf-Rayet stars in the Tarantula does not exclude the binary channel 
since low-mass stripped stars would be unlikely to exhibit a Wolf-Rayet spectral appearance 
\cite{Gotberg18}.

Initially very close binaries may merge on the main sequence, prior to following a relatively 
conventional evolution, albeit with unusually high rotation rates, which would lead to increased 
luminosities and potentially evolve blueward off the main sequence \cite{Brott11, Kohler15}. 
Extremely rapid rotation in some VFTS OB stars favours close binary evolution or stellar mergers. 
Very massive stars in the Tarantula up to $\sim 300 M_{\odot}$ are expected to lead to 30--50 
$M_{\odot}$ CO cores and black hole fates, unlikely to produce any associated supernova 
\cite{Yusof13}, such that a subset of binary VMS are plausible progenitors of LIGO black hole 
binary mergers, although their exact fate crucially depends on their mass-loss properties, which 
remain uncertain.



\begin{figure}[htbp]
\centering
\includegraphics[width=9 cm]{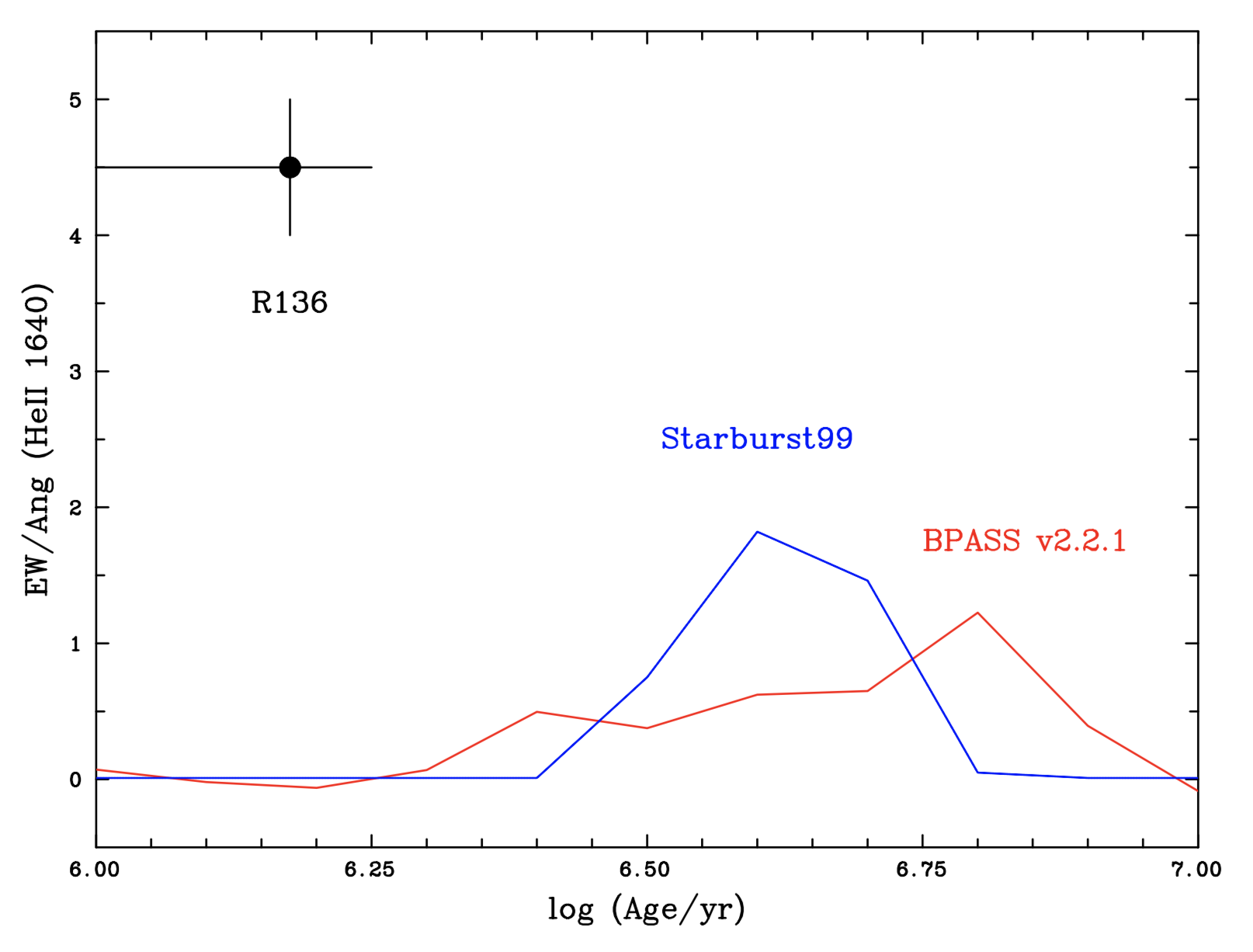}
\caption{Comparison between observed He\,{\sc ii} $\lambda$1640 emission equivalent widths in R136 \cite{Crowther16} versus predicted emission from BPASS (v.2.2.1, red) and Starburst99 (blue) population synthesis models (absorption lines are shown as negative values).}\label{1640}
\end{figure}

\section{Integrated properties and comparison with star-forming regions, near and far}

The Tarantula Nebula would subtend little more than one arcsec if it were located at a distance of 
50 Mpc, and  so provides us with a unique opportunity to compare the individual spatially-resolved 
properties of an intensively star-forming region and its aggregate characteristics. Using 
integrated H$\alpha$ observations of the Tarantula Nebula \cite{Pellegrini10}, an age of $\sim$3.5 
Myr would be inferred from a comparison between the inferred H$\alpha$ equivalent width of 1100\AA\ 
and population synthesis models for a coeval population at LMC metallicity \cite{Doran13}. This is 
in reasonable agreement with the typical age of massive stars, albeit failing to reflect the 
complexity in its star-formation history (recall Fig.~\ref{HRD}).

An analysis of the integrated UV spectrum of NGC~2070 supports a young ($\leq$3 Myr) starburst 
episode \cite{Vacca95}, while the high spatial resolution of HST/STIS has permitted a comparison 
between the individual and integrated UV spectroscopic appearance of the central R136 cluster 
\cite{Crowther16}. Very massive stars contributed a significant fraction of its far UV continuum 
flux, and completely dominate the strong, broad He\,{\sc ii} $\lambda$1640 emission. The integrated 
UV spectroscopic appearance of R136 closely resembles some star clusters in star-forming galaxies 
at Mpc distances, such as NGC5253-5 \cite{Smith16}, suggesting the presence of VMS in these other 
young massive clusters. From comparison with the predictions of standard population synthesis 
models, both Starburst99 \cite{Levesque12} and BPASS \cite{Eldridge17} models fail to predict any 
significant emission prior to the conventional Wolf-Rayet phase (Fig.~\ref{1640}), owing to the use 
of inadequate wind theory for VMS \cite{Grafener11, Bestenlehner19a}. Consequently, neither 
Starburst99 nor BPASS accounts for the powerful winds of very massive stars in R136, and the 
adopted mass function follows a Salpeter slope, rather than the top heavy IMF identified by 
\cite{Schneider18a} for the Tarantula region as a whole.


\begin{table}[htbp]
\caption{Top ten stellar systems contributing to the Lyman continuum output of the Tarantula Nebula, comprising very massive early O stars and WN5 stars, and 
classical Wolf-Rayet stars, updated from \cite{Doran13}.}\label{monsters} 
\centering
\begin{tabular}{lcccl}
\toprule
\textbf{Star (alias)}	& \textbf{Sp Type}	& \textbf{$\log L/L_{\odot}$} & \textbf{$\log$ N(LyC)} & \textbf{Reference} \\
\midrule
R136a1 (BAT99-108)    &. WN5h              & 6.8               &50.6 & \cite{Bestenlehner19b} \\ 
Mk\,34 (BAT99-116)        & WN5h+WN5h   & 6.4+6.4       &50.6 & \cite{Tehrani19} \\  
R136a2 (BAT99-109)    & WN5h               & 6.7               &50.5 & \cite{Bestenlehner19b} \\ %
R144 (BAT99-118)      & WN5-6+WN6-7 & 6.7             & 50.5  & \cite{Sana13, Hainich14}\\ 
R136a3 (BAT99-106)    & WN5h                & 6.6               & 50.5 & \cite{Bestenlehner19b} \\ 
Mk\,49 (BAT99-98)         & WN6(h)             & 6.7             &50.5 & \cite{Hainich14} \\ 
R145 (BAT99-119)         & WN6h+O3.5If/WN7 & 6.3+6.3 & 50.4 & \cite{Shenar19} \\ 
Mk\,42 (BAT99-105)        & O2\,If                 & 6.6               &50.4 & \cite{Bestenlehner14} \\ 
VFTS 682                       & WN5h               & 6.5               & 50.4 & \cite{Bestenlehner14}  \\ 
R136c (BAT99-112)      & WN5h+?            &6.6               & 50.4 & \cite{Bestenlehner14}  \\ 
\bottomrule
\end{tabular}
\end{table}

\begin{figure}[htbp]
\centering
\includegraphics[width=12 cm]{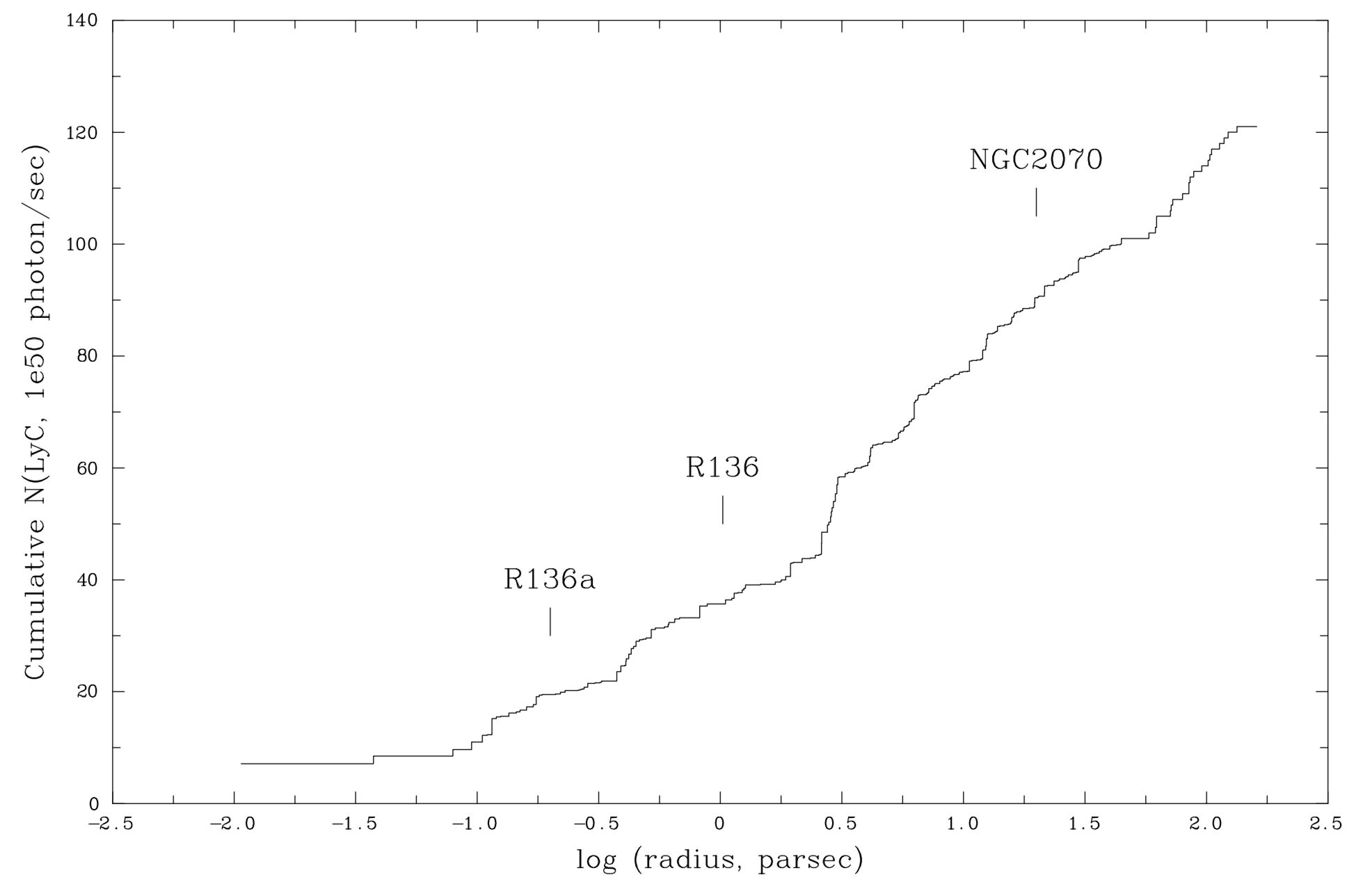}
\caption{Cumulative ionizing output (10$^{50}$ ph/s) from spectroscopically classified early-type stars in the Tarantula, obtained from VFTS 
\cite{Bestenlehner14, SabinSanjulian17, RamirezAgudelo17}, VLT/MUSE \cite{Castro19}, HST/STIS \cite{Bestenlehner19b} and literature results \cite{Hainich14}, updated from \cite{Doran13}. Specific regions within 30 Dor are indicated from Table~\ref{scale}}
\label{Q0}
\end{figure}

An estimate of the cumulative ionizing and mechanical feedback from massive stars within the 
Tarantula has revealed a major contribution from VMS towards the collective ionizing output and a 
dominant role from WR stars to the mechanical feedback \cite{Doran13}. However this analysis relied 
on calibrations and estimates of spectral types for a significant subset of the massive star 
content, so we are able to provide updates from recent spectroscopic observations (e.g. VLT/MUSE, 
HST/STIS) and analyses. We present the updated cumulative ionizing output from 1170 massive stars 
in the Tarantula Nebula in Figure~\ref{Q0}, indicating a total Lyman continuum ionizing output of 
1.2$\times 10^{52}$ ph\,s$^{-1}$ within 150 pc of R136a. A quarter of the total ionizing radiation 
originates from the R136a cluster, while members of NGC~2070 produce three quarters of the global 
feedback. Ten systems alone, listed in Table~\ref{monsters}, collectively contribute a quarter of 
the ionizing budget of the Tarantula Nebula, comprising main sequence and classical WR stars, plus 
early O supergiants. Indeed, half of the global ionizing output originates from 40 early O 
stars, main sequence WN stars and classical WR stars, with 1130 stars contributing the remaining 
50\%. The collective bolometric luminosity of these stars is $10^{8.4} L_{\odot}$, of which the 40 UV-bright stars contribute $10^{8.0} L_{\odot}$.

Recalling Table~\ref{GHII},  the highest mass stars and evolved high mass stars in other giant 
H\,{\sc ii} regions in the Local Group dominate their radiative and mechanical feedback, including 
NGC~3372 (Carina Nebula) in the Milky Way, N206 in the LMC and NGC~346 in the SMC. By way of 
example, Smith \cite{Smith06} established that only a handful of early O-type stars and H-rich WN 
stars contribute the majority of the Lyman continuum flux of the Carina Nebula, while $\eta$ Car, 
four Wolf-Rayet stars and two early O supergiants completely dominate the stellar mechanical 
luminosity. The central ionizing cluster of the Galactic NGC~3603 star-forming region is host to a 
stellar content analogous to R136a, including a number of early O stars, nitrogen-sequence 
Wolf-Rayet stars \cite{Moffat02, Melena08}. Weak main-sequence wind properties of metal-poor 
massive stars conspire to even fewer massive stars (HD~5980, Sk\,80) dominating the cumulative 
stellar feedback in NGC~346. Similar conclusions were reached by Ramachandran et al. 
\cite{Ramachandran19} for the supergiant shell in the wing of the SMC.

\begin{figure}[htbp]
\centering
\includegraphics[width=10 cm]{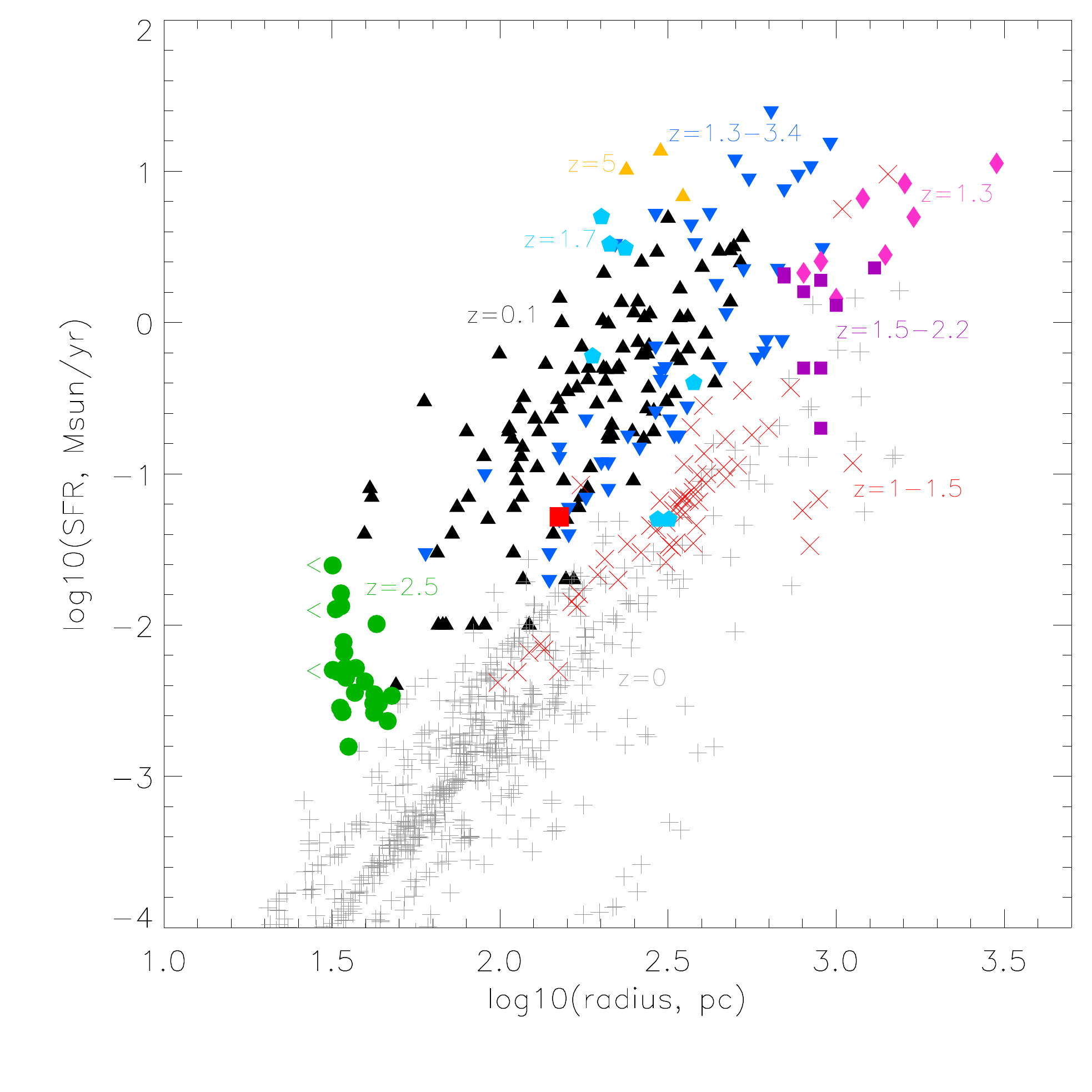}
\caption{Comparison between the integrated star-formation rate versus size of the Tarantula (filled red square) and star-forming knots from galaxies spanning a range of redshifts, adapted from \cite{Johnson17}.}
\label{SFR}
\end{figure}


Although the Tarantula Nebula is the most extreme giant H\,{\sc ii} region in the Local Group, how 
does it rank against star-forming regions of galaxies in the near universe or knots at high 
redshift? Fig~\ref{SFR} compares the star-formation rate versus size of 
regions spanning $z$=0 to 3.4, adapted from \cite{Johnson17}, indicating that the Tarantula (red 
square) is forming stars more vigorously than typical low-redshift counterparts, resembling some 
star-forming regions at high redshift. Indeed, 80\% of the cumulative ionizing radiation originates 
from NGC~2070, such that this region corresponds closely with typical clumps in the lensed galaxy 
SDSS J1110+6459 at $z$=2.5 (green circles).

\begin{figure}[htbp]
\centering
\includegraphics[width=10 cm]{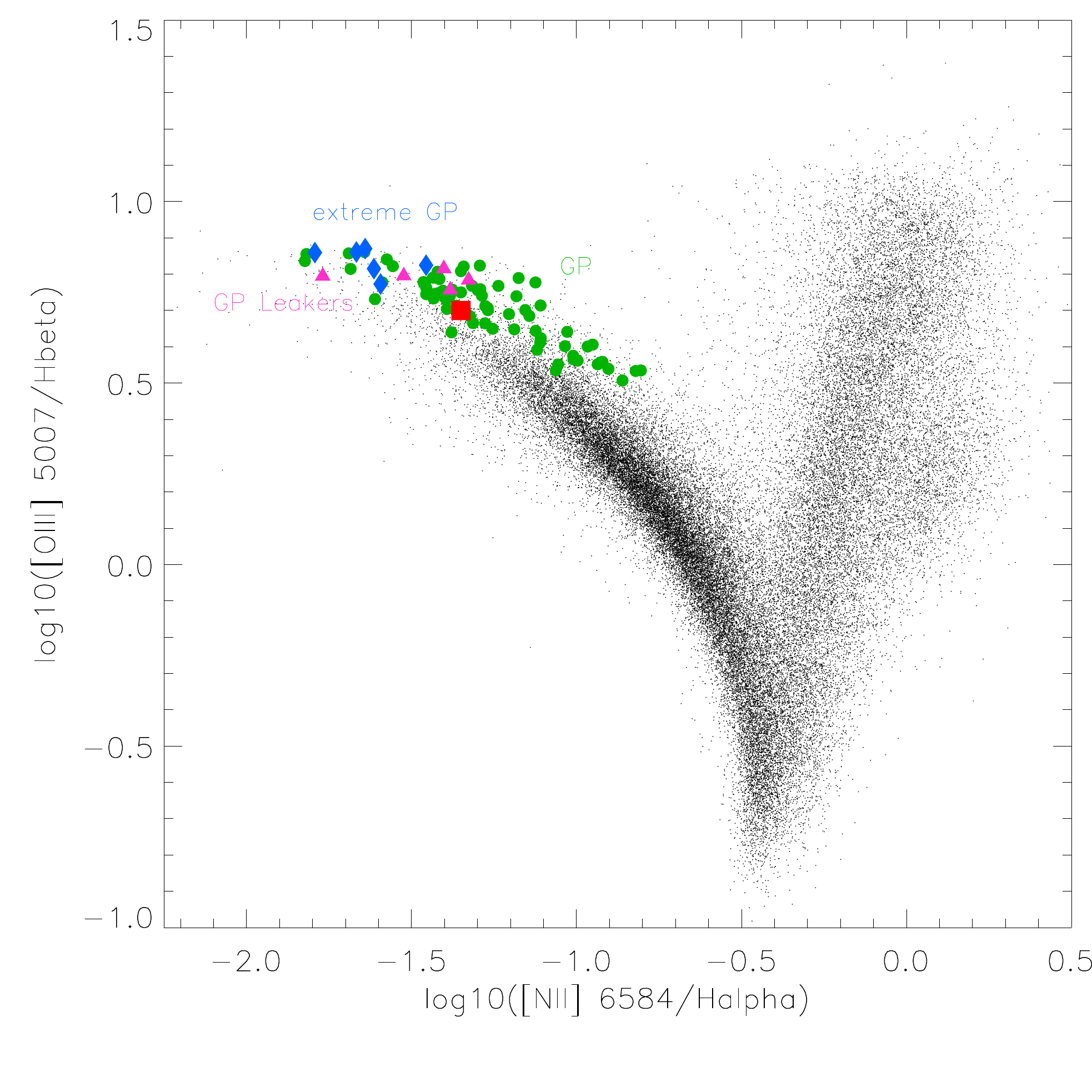}
\caption{BPT diagram illustrating the similarity in integrated strengths between the Tarantula Nebula (red square), Green Pea (green circles), extreme Green Peas (blue diamonds), Lyman-continuum leaking Green Peak (pink triangles), updated from \cite{Micheva17}, plus SDSS star-forming galaxies.}
\label{BPT}
\end{figure}
\
In addition to its unusually high star formation rate, the Tarantula also possesses high ionization 
parameter nebular properties with respect to star-forming galaxies in the local universe. Fig~\ref{BPT} presents a Baldwin, Philipps \& Terlevich \cite[BPT]{Baldwin81} diagnostic 
diagram of Sloan Digital Sky Survey (SDSS) galaxies, in which the Tarantula (red square) has been 
indicated, along with Green Pea galaxies from Micheva et al. \cite{Micheva17} which are 
low-metallicity, intensively star-forming galaxies exhibiting unusually strong [O\,{\sc iii}] 
$\lambda$5007 emission. Steidel et al. \cite{Steidel14} showed that $z$=2--3 star forming galaxies 
share similar extreme nebular properties, and a subset of Green Pea galaxies have been established 
as Lyman continuum leakers \cite{Izotov16a, Izotov16b}. Focusing again on NGC~2070, this sits 
amongst the extreme Green Pea galaxies in the BPT diagram.


\begin{figure}[htbp]
\centering
\includegraphics[width=10 cm]{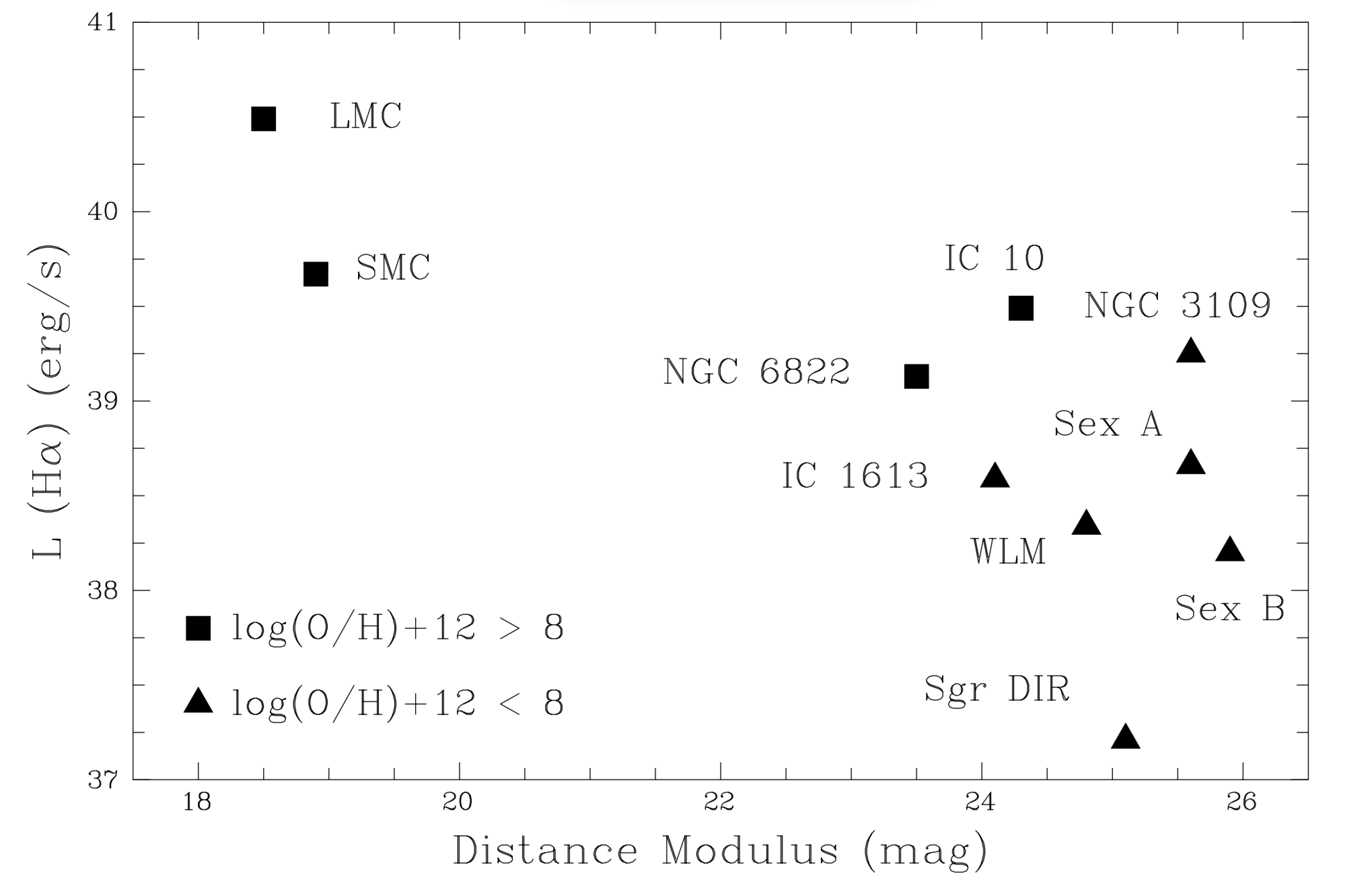}
\caption{Comparison between present-day star formation rates, as measured by H$\alpha$ luminosity \cite{Kennicutt08}, distance modulus (mag), and oxygen metal content (squares: $\geq$20\% of solar value, triangles: $<$20\% of solar value, for Local Group dwarf galaxies. Metal-poor galaxies possess low star-formation rates, so host small
numbers of OB stars, and these are $\geq$6 magnitudes fainter than Magellanic Cloud counterparts.}\label{SFR}
\end{figure}

\section{Summary and outlook}

Recent comprehensive spectroscopic and imaging surveys have revealed that the Tarantula Nebula 
hosts the most exceptional massive star population within the Local Group of galaxies, including 
the most massive stars identified to date, the fastest rotating early-type stars, and the X-ray 
brightest colliding wind system. In particular, the VFTS survey has revealed an excess of massive 
stars with respect to a Salpeter IMF \cite{Schneider18a} and added support from previous results 
for the importance of close binary evolution in the evolution of massive stars \cite{Sana13}. As 
such, the Tarantula Nebula is the closest analogue to the Hubble Deep Field for the community 
interested in the evolution of massive stars since its richness provides us with a huge breadth of 
extreme stars.

The integrated appearance of R136 resembles young extragalactic star clusters, while the integrated 
nebular properties of NGC~2070 is analogous to extreme Green Pea galaxies at low redshift and 
star-forming knots in high-redshift galaxies. Typical metallicities of Green Pea galaxies are lower 
than the LMC, as measured from oxygen nebular lines, while the oxygen content of high-$z$ star 
forming galaxies tends to be similar to those of the Magellanic Clouds. However, $\alpha$/Fe 
abundances of high redshift galaxies are likely to be higher than young populations in the Milky 
Way, such that winds from OB stars at high redshift are anticipated to be weaker than in the LMC or 
SMC \cite{Steidel16}.

The LMC metallicity is only a factor of two below that of the Solar neighbourhood \cite{Tsamis05}, 
so ideally we would like to supplement the extensive survey of the Tarantula Nebula with 
counterparts at significantly lower metallicity. The SMC (1/5 solar) represents our best 
opportunity to study the formation and evolution of massive stars at a metallicity significantly 
below that of the LMC. Alas, it does not host as rich a massive star-forming region as the 
Tarantula, but cumulatively does host a substantial number of O stars so is key towards our 
improved understanding of massive stars at low metallicity, especially as it can be studied in 
exquisite detail with current instrumentation.

\begin{figure}[htbp]
\centering
\includegraphics[width=10 cm]{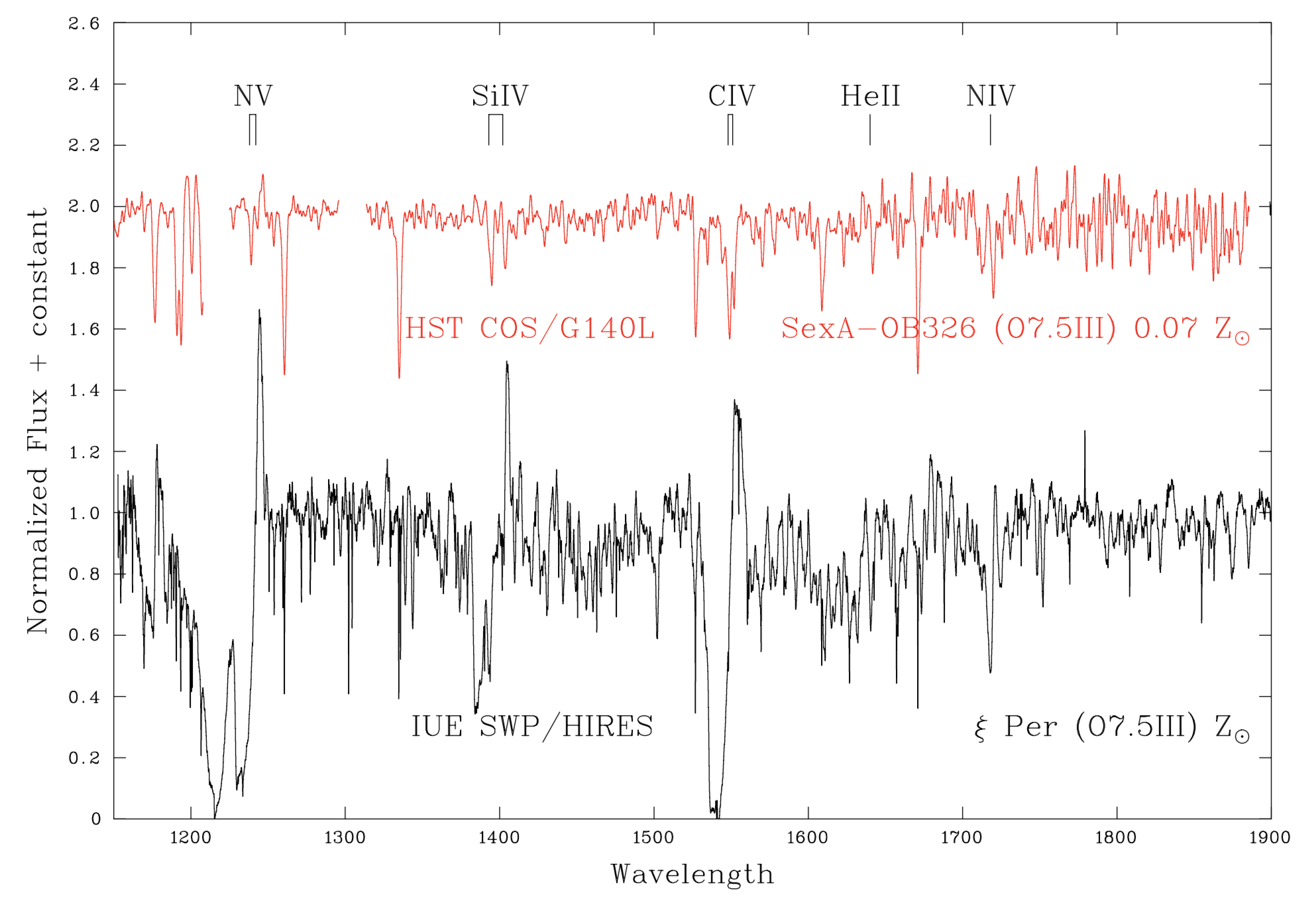}
\caption{Comparison between far UV spectroscopy of mid O giants in metal-rich \cite{Walborn85} and metal-deficient \cite{Garcia19} environments, illustrating the extreme differences in wind features (e.g. N\.,{\sc v} $\lambda$1240, Si\,{\sc iv} $\lambda$1400, C\,{\sc iv} $\lambda$1550) and the iron forest (Fe\,{\sc iv-v}).}\label{o7giant}
\end{figure}

Since we need to look beyond the SMC to study metal-poor counterparts to the Tarantula Nebula, 
Fig~\ref{SFR} compares the star-formation rate, metallicity, and distance modulus of Local Group 
dwarf galaxies. Rates of star formation in metal-poor ($\leq$20\% of solar oxygen content) galaxies 
are significantly lower than the Magellanic Clouds, so there are no rich metal-poor massive star 
populations elsewhere in the Local Group, and those few that are present are much more distant. The 
absence of nearby metal-poor counterparts to the Tarantula implies that we currently have to rely 
on the interpretation of integrated populations in order to understand massive stellar evolution at 
low metallicity, notably extremely metal-poor dwarf star-forming galaxies I\,Zw~18 and SBS-0335.

In order to test predictions of the metallicity dependence of massive star winds, it is necessary 
to measure mass-loss properties across a wide range of metallicities. Although theory has been 
qualitatively supported from the observed wind properties of Milky Way, LMC and SMC early-type 
stars \cite{Mokiem07}, some issues remain, including weak winds in low luminosity OB stars. Our 
only opportunity to study individual massive stars below 1/10th of the solar oxygen content is to 
observe O stars in Sextans A and B with 7\% solar \cite{vanZee06} or the Sagittarius Dwarf 
Irregular Galaxy (SgrDIR) with 5\% solar \cite{Saviane02}. Stellar winds from early-type stars at 
such low metallicities are anticipated to be much weaker than in metal-rich populations, which has 
been confirmed by UV spectroscopy. By way of example, Fig.~\ref{o7giant} compares the far UV 
spectrum of $\xi$ Per (O7.5\,{\sc iii}) with a counterpart in Sextans A \cite{Garcia19}, revealing 
negligible wind signatures in the latter (e.g. C\,{\sc iv} $\lambda$1550).

\vspace{6pt} 




\funding{This research received no external funding}

\acknowledgments{This review is dedicated to the memory of Nolan Walborn, from whom the author learnt a great deal about the Tarantula Nebula. Thanks to Leisa Townsley and Patrick Broos for access to T-ReX point source results prior to publication, Miriam Garcia for the UV spectrum of Sextans A OB\,326, Selma De Mink for the VFTS pie chart, Heloise Stevance for the BPASS predictions, and Joachim Bestenlehner for converting $Q$ wind density results into mass-loss rates. Feedback
from Fabian Schneider, Chris Evans, Joachim Bestenlehner, Andy Pollock, Roberta Humphreys and external referees on an earlier draft is greatly appreciated.}

\conflictsofinterest{The author declares no conflict of interest.} 





\reftitle{References}


\externalbibliography{yes}
\bibliography{tarantula}

\begin{thebibliography}{-------}
\providecommand{\natexlab}[1]{#1}

\bibitem[{Kennicutt}(1998)]{Kennicutt98}
{Kennicutt}, Robert~C., J.
\newblock {Star Formation in Galaxies Along the Hubble Sequence}.
\newblock {\em \araa} {\bf 1998}, {\em 36},~189--232,
  \href{http://xxx.lanl.gov/abs/astro-ph/9807187}{{\normalfont
  [arXiv:astro-ph/astro-ph/9807187]}}.
\newblock
  doi:{\changeurlcolor{black}\href{https://doi.org/10.1146/annurev.astro.36.1.189}{\detokenize{10.1146/annurev.astro.36.1.189}}}.

\bibitem[{Evans} \em{et~al.}(2011){Evans}, {Taylor}, {H{\'e}nault-Brunet},
  {Sana}, {de Koter}, {Sim{\'o}n-D{\'\i}az}, {Carraro}, {Bagnoli}, {Bastian},
  {Bestenlehner}, {Bonanos}, {Bressert}, {Brott}, {Campbell}, {Cantiello},
  {Clark}, {Costa}, {Crowther}, {de Mink}, {Doran}, {Dufton}, {Dunstall},
  {Friedrich}, {Garcia}, {Gieles}, {Gr{\"a}fener}, {Herrero}, {Howarth},
  {Izzard}, {Langer}, {Lennon}, {Ma{\'\i}z Apell{\'a}niz}, {Markova},
  {Najarro}, {Puls}, {Ramirez}, {Sab{\'\i}n-Sanjuli{\'a}n}, {Smartt}, {Stroud},
  {van Loon}, {Vink}, and {Walborn}]{Evans11}
{Evans}, C.J.; {Taylor}, W.D.; {H{\'e}nault-Brunet}, V.; {Sana}, H.; {de
  Koter}, A.; {Sim{\'o}n-D{\'\i}az}, S.; {Carraro}, G.; {Bagnoli}, T.;
  {Bastian}, N.; {Bestenlehner}, J.M. et al.
\newblock {The VLT-FLAMES Tarantula Survey. I. Introduction and observational
  overview}.
\newblock {\em \aap} {\bf 2011}, {\em 530},~A108,
  \href{http://xxx.lanl.gov/abs/1103.5386}{{\normalfont
  [arXiv:astro-ph.SR/1103.5386]}}.
\newblock
doi:{\changeurlcolor{black}\href{https://doi.org/10.1051/0004-6361/201116782}{\detokenize{10.1051/0004-6361/201116782}}}.


\bibitem[{Tsamis} and {P{\'e}quignot}(2005)]{Tsamis05}
{Tsamis}, Y.G.; {P{\'e}quignot}, D.
\newblock {A photoionization-modelling study of 30 Doradus: the case for
  small-scale chemical inhomogeneity}.
\newblock {\em \mnras} {\bf 2005}, {\em 364},~687--704,
  \href{http://xxx.lanl.gov/abs/astro-ph/0509463}{{\normalfont
  [astro-ph/0509463]}}.
\newblock
doi:{\changeurlcolor{black}\href{https://doi.org/10.1111/j.1365-2966.2005.09595.x}{\detokenize{10.1111/j.1365-2966.2005.09595.x}}}.


\bibitem[{Baldwin} \em{et~al.}(1991){Baldwin}, {Ferland}, {Martin}, {Corbin},
  {Cota}, {Peterson}, and {Slettebak}]{Baldwin91}
{Baldwin}, J.A.; {Ferland}, G.J.; {Martin}, P.G.; {Corbin}, M.R.; {Cota}, S.A.;
  {Peterson}, B.M.; {Slettebak}, A.
\newblock {Physical conditions in the Orion Nebula and an assessment of its
  helium abundance}.
\newblock {\em \apj} {\bf 1991}, {\em 374},~580--609.
\newblock
  doi:{\changeurlcolor{black}\href{https://doi.org/10.1086/170146}{\detokenize{10.1086/170146}}}.


\bibitem[{Crowther} \em{et~al.}(2016){Crowther}, {Caballero-Nieves},
  {Bostroem}, {Ma{\'\i}z Apell{\'a}niz}, {Schneider}, {Walborn}, {Angus},
  {Brott}, {Bonanos}, {de Koter}, {de Mink}, {Evans}, {Gr{\"a}fener},
  {Herrero}, {Howarth}, {Langer}, {Lennon}, {Puls}, {Sana}, and
  {Vink}]{Crowther16}
{Crowther}, P.A.; {Caballero-Nieves}, S.M.; {Bostroem}, K.A.; {Ma{\'\i}z
  Apell{\'a}niz}, J.; {Schneider}, F.R.N.; {Walborn}, N.R.; {Angus}, C.R.;
  {Brott}, I.; {Bonanos}, A.; {de Koter}, A. et al.
\newblock {The R136 star cluster dissected with Hubble Space Telescope/STIS. I.
  Far-ultraviolet spectroscopic census and the origin of He II
  {\ensuremath{\lambda}}1640 in young star clusters}.
\newblock {\em \mnras} {\bf 2016}, {\em 458},~624--659,
  \href{http://xxx.lanl.gov/abs/1603.04994}{{\normalfont
  [arXiv:astro-ph.SR/1603.04994]}}.
\newblock
  doi:{\changeurlcolor{black}\href{https://doi.org/10.1093/mnras/stw273}{\detokenize{10.1093/mnras/stw273}}}.

  
\bibitem[{Britavskiy} \em{et~al.}(2019){Britavskiy}, {Lennon}, {Patrick},
  {Evans}, {Herrero}, {Langer}, {van Loon}, {Clark}, {Schneider}, {Almeida},
  {Sana}, {de Koter}, and {Taylor}]{Britavskiy19}
{Britavskiy}, N.; {Lennon}, D.J.; {Patrick}, L.R.; {Evans}, C.J.; {Herrero},
  A.; {Langer}, N.; {van Loon}, J.T.; {Clark}, J.S.; {Schneider}, F.R.N.;
  {Almeida}, L.A. et al.
\newblock {The VLT-FLAMES Tarantula Survey. XXX. Red stragglers in the clusters
  Hodge 301 and SL 639}.
\newblock {\em \aap} {\bf 2019}, {\em 624},~A128,
  \href{http://xxx.lanl.gov/abs/1902.09891}{{\normalfont
  [arXiv:astro-ph.SR/1902.09891]}}.
\newblock
  doi:{\changeurlcolor{black}\href{https://doi.org/10.1051/0004-6361/201834564}{\detokenize{10.1051/0004-6361/201834564}}}.

\bibitem[{Walborn}(1991)]{Walborn91}
{Walborn}, N.R.
\newblock {The Starburst Region 30 Doradus}.
\newblock  The Magellanic Clouds; {Haynes}, R.; {Milne}, D., Eds.,  1991, Vol.
  148, {\em IAU Symposium}, p. 145.

\bibitem[{Sabbi} \em{et~al.}(2013){Sabbi}, {Anderson}, {Lennon}, {van der
  Marel}, {Aloisi}, {Boyer}, {Cignoni}, {de Marchi}, {de Mink}, {Evans},
  {Gallagher}, {Gordon}, {Gouliermis}, {Grebel}, {Koekemoer}, {Larsen},
  {Panagia}, {Ryon}, {Smith}, {Tosi}, and {Zaritsky}]{Sabbi13}
{Sabbi}, E.; {Anderson}, J.; {Lennon}, D.J.; {van der Marel}, R.P.; {Aloisi},
  A.; {Boyer}, M.L.; {Cignoni}, M.; {de Marchi}, G.; {de Mink}, S.E.; {Evans},
  C.J. et al.
\newblock {Hubble Tarantula Treasury Project: Unraveling Tarantula's Web. I.
  Observational Overview and First Results}.
\newblock {\em \aj} {\bf 2013}, {\em 146},~53,
  \href{http://xxx.lanl.gov/abs/1304.6747}{{\normalfont
  [arXiv:astro-ph.GA/1304.6747]}}.
\newblock
  doi:{\changeurlcolor{black}\href{https://doi.org/10.1088/0004-6256/146/3/53}{\detokenize{10.1088/0004-6256/146/3/53}}}.

\bibitem[{Crowther} \em{et~al.}(2010){Crowther}, {Schnurr}, {Hirschi}, {Yusof},
  {Parker}, {Goodwin}, and {Kassim}]{Crowther10}
{Crowther}, P.A.; {Schnurr}, O.; {Hirschi}, R.; {Yusof}, N.; {Parker}, R.J.;
  {Goodwin}, S.P.; {Kassim}, H.A.
\newblock {The R136 star cluster hosts several stars whose individual masses
  greatly exceed the accepted 150M$_{solar}$ stellar mass limit}.
\newblock {\em \mnras} {\bf 2010}, {\em 408},~731--751,
  \href{http://xxx.lanl.gov/abs/1007.3284}{{\normalfont
  [arXiv:astro-ph.SR/1007.3284]}}.
\newblock
  doi:{\changeurlcolor{black}\href{https://doi.org/10.1111/j.1365-2966.2010.17167.x}{\detokenize{10.1111/j.1365-2966.2010.17167.x}}}.

\bibitem[{Castro} \em{et~al.}(2018){Castro}, {Crowther}, {Evans}, {Mackey},
  {Castro-Rodriguez}, {Vink}, {Melnick}, and {Selman}]{Castro18}
{Castro}, N.; {Crowther}, P.A.; {Evans}, C.J.; {Mackey}, J.;
  {Castro-Rodriguez}, N.; {Vink}, J.S.; {Melnick}, J.; {Selman}, F.
\newblock {Mapping the core of the Tarantula Nebula with VLT-MUSE. I. Spectral
  and nebular content around R136}.
\newblock {\em \aap} {\bf 2018}, {\em 614},~A147,
  \href{http://xxx.lanl.gov/abs/1802.01597}{{\normalfont [1802.01597]}}.
\newblock
  doi:{\changeurlcolor{black}\href{https://doi.org/10.1051/0004-6361/201732084}{\detokenize{10.1051/0004-6361/201732084}}}.

\bibitem[{Castro} \em{et~al.}(in prep.){Castro}, {Crowther}, and
  {Evans}]{Castro19}
{Castro}, N.; {Crowther}, P.A.; {Evans}, C.J. et al.
\newblock {Mapping the core of the Tarantula Nebula with VLT-MUSE. III. The
  spectroscopic Hertzsprung-Russell diagram of OB stars in NGC2070}.
\newblock {\em \aap} {\bf in prep.}

\bibitem[{Doran} \em{et~al.}(2013){Doran}, {Crowther}, {de Koter}, {Evans},
  {McEvoy}, {Walborn}, {Bastian}, {Bestenlehner}, {Gr{\"a}fener}, {Herrero},
  {K{\"o}hler}, {Ma{\'{\i}}z Apell{\'a}niz}, {Najarro}, {Puls}, {Sana},
  {Schneider}, {Taylor}, {van Loon}, and {Vink}]{Doran13}
{Doran}, E.I.; {Crowther}, P.A.; {de Koter}, A.; {Evans}, C.J.; {McEvoy}, C.;
  {Walborn}, N.R.; {Bastian}, N.; {Bestenlehner}, J.M.; {Gr{\"a}fener}, G.;
  {Herrero}, A. et al.
\newblock {The VLT-FLAMES Tarantula Survey. XI. A census of the hot luminous
  stars and their feedback in 30 Doradus}.
\newblock {\em \aap} {\bf 2013}, {\em 558},~A134,
  \href{http://xxx.lanl.gov/abs/1308.3412}{{\normalfont
  [arXiv:astro-ph.SR/1308.3412]}}.
\newblock
  doi:{\changeurlcolor{black}\href{https://doi.org/10.1051/0004-6361/201321824}{\detokenize{10.1051/0004-6361/201321824}}}.

\bibitem[{De Marchi} \em{et~al.}(2011){De Marchi}, {Paresce}, {Panagia},
  {Beccari}, {Spezzi}, {Sirianni}, {Andersen}, {Mutchler}, {Balick}, {Dopita},
  {Frogel}, {Whitmore}, {Bond}, {Calzetti}, {Carollo}, {Disney}, {Hall},
  {Holtzman}, {Kimble}, {McCarthy}, {O'Connell}, {Saha}, {Silk}, {Trauger},
  {Walker}, {Windhorst}, and {Young}]{DeMarchi11}
{De Marchi}, G.; {Paresce}, F.; {Panagia}, N.; {Beccari}, G.; {Spezzi}, L.;
  {Sirianni}, M.; {Andersen}, M.; {Mutchler}, M.; {Balick}, B.; {Dopita}, M.A. et al.
\newblock {Star Formation in 30 Doradus}.
\newblock {\em \apj} {\bf 2011}, {\em 739},~27,
  \href{http://xxx.lanl.gov/abs/1106.2801}{{\normalfont
  [arXiv:astro-ph.SR/1106.2801]}}.
\newblock
  doi:{\changeurlcolor{black}\href{https://doi.org/10.1088/0004-637X/739/1/27}{\detokenize{10.1088/0004-637X/739/1/27}}}.

\bibitem[{Sabbi} \em{et~al.}(2016){Sabbi}, {Lennon}, {Anderson}, {Cignoni},
  {van der Marel}, {Zaritsky}, {De Marchi}, {Panagia}, {Gouliermis}, {Grebel},
  {Gallagher}, {Smith}, {Sana}, {Aloisi}, {Tosi}, {Evans}, {Arab}, {Boyer}, {de
  Mink}, {Gordon}, {Koekemoer}, {Larsen}, {Ryon}, and {Zeidler}]{Sabbi16}
{Sabbi}, E.; {Lennon}, D.J.; {Anderson}, J.; {Cignoni}, M.; {van der Marel},
  R.P.; {Zaritsky}, D.; {De Marchi}, G.; {Panagia}, N.; {Gouliermis}, D.A.;
  {Grebel}, E.K.; {Gallagher}, J.~S. et al.
\newblock {Hubble Tarantula Treasury Project. III. Photometric Catalog and
  Resulting Constraints on the Progression of Star Formation in the 30 Doradus
  Region}.
\newblock {\em \apjs} {\bf 2016}, {\em 222},~11,
  \href{http://xxx.lanl.gov/abs/1511.06021}{{\normalfont
  [arXiv:astro-ph.GA/1511.06021]}}.
\newblock
  doi:{\changeurlcolor{black}\href{https://doi.org/10.3847/0067-0049/222/1/11}{\detokenize{10.3847/0067-0049/222/1/11}}}.

\bibitem[{Schneider} \em{et~al.}(2018){Schneider}, {Ram{\'\i}rez-Agudelo},
  {Tramper}, {Bestenlehner}, {Castro}, {Sana}, {Evans},
  {Sab{\'\i}n-Sanjuli{\'a}n}, {Sim{\'o}n-D{\'\i}az}, {Langer}, {Fossati},
  {Gr{\"a}fener}, {Crowther}, {de Mink}, {de Koter}, {Gieles}, {Herrero},
  {Izzard}, {Kalari}, {Klessen}, {Lennon}, {Mahy}, {Ma{\'\i}z Apell{\'a}niz},
  {Markova}, {van Loon}, {Vink}, and {Walborn}]{Schneider18b}
{Schneider}, F.R.N.; {Ram{\'\i}rez-Agudelo}, O.H.; {Tramper}, F.;
  {Bestenlehner}, J.M.; {Castro}, N.; {Sana}, H.; {Evans}, C.J.;
  {Sab{\'\i}n-Sanjuli{\'a}n}, C.; {Sim{\'o}n-D{\'\i}az}, S.; {Langer}, N. et al.
\newblock {The VLT-FLAMES Tarantula Survey. XXIX. Massive star formation in the
  local 30 Doradus starburst}.
\newblock {\em \aap} {\bf 2018}, {\em 618},~A73,
  \href{http://xxx.lanl.gov/abs/1807.03821}{{\normalfont
  [arXiv:astro-ph.SR/1807.03821]}}.
\newblock
  doi:{\changeurlcolor{black}\href{https://doi.org/10.1051/0004-6361/201833433}{\detokenize{10.1051/0004-6361/201833433}}}.

\bibitem[{Wright} \em{et~al.}(2016){Wright}, {Bouy}, {Drew}, {Sarro}, {Bertin},
  {Cuillandre}, and {Barrado}]{Wright16}
{Wright}, N.J.; {Bouy}, H.; {Drew}, J.E.; {Sarro}, L.M.; {Bertin}, E.;
  {Cuillandre}, J.C.; {Barrado}, D.
\newblock {Cygnus OB2 DANCe: A high-precision proper motion study of the Cygnus
  OB2 association}.
\newblock {\em \mnras} {\bf 2016}, {\em 460},~2593--2610,
  \href{http://xxx.lanl.gov/abs/1605.03583}{{\normalfont
  [arXiv:astro-ph.SR/1605.03583]}}.
\newblock
  doi:{\changeurlcolor{black}\href{https://doi.org/10.1093/mnras/stw1148}{\detokenize{10.1093/mnras/stw1148}}}.

  
\bibitem[{Walborn}(2002)]{Walborn02}
{Walborn}, N.R.
\newblock {The Pillars of the Second Generation}.
\newblock  Hot Star Workshop III: The Earliest Phases of Massive Star Birth;
  {Crowther}, P.A., Ed.,  2002, Vol. 267, {\em Astronomical Society of the
  Pacific Conference Series}, p. 111.


\bibitem[{Walborn} \em{et~al.}(2013){Walborn}, {Barb{\'a}}, and
  {Sewi{\l}o}]{Walborn13}
{Walborn}, N.R.; {Barb{\'a}}, R.H.; {Sewi{\l}o}, M.M.
\newblock {The Top 10 Spitzer Young Stellar Objects in 30 Doradus}.
\newblock {\em \aj} {\bf 2013}, {\em 145},~98.
\newblock
  doi:{\changeurlcolor{black}\href{https://doi.org/10.1088/0004-6256/145/4/98}{\detokenize{10.1088/0004-6256/145/4/98}}}.


\bibitem[{Indebetouw} \em{et~al.}(2013){Indebetouw}, {Brogan}, {Chen}, {Leroy},
  {Johnson}, {Muller}, {Madden}, {Cormier}, {Galliano}, {Hughes}, {Hunter},
  {Kawamura}, {Kepley}, {Lebouteiller}, {Meixner}, {Oliveira}, {Onishi}, and
  {Vasyunina}]{Indebetouw13}
{Indebetouw}, R.; {Brogan}, C.; {Chen}, C.H.R.; {Leroy}, A.; {Johnson}, K.;
  {Muller}, E.; {Madden}, S.; {Cormier}, D.; {Galliano}, F.; {Hughes}, A. et al.
\newblock {ALMA Resolves 30 Doradus: Sub-parsec Molecular Cloud Structure near
  the Closest Super Star Cluster}.
\newblock {\em \apj} {\bf 2013}, {\em 774},~73,
  \href{http://xxx.lanl.gov/abs/1307.3680}{{\normalfont [1307.3680]}}.
\newblock
  doi:{\changeurlcolor{black}\href{https://doi.org/10.1088/0004-637X/774/1/73}{\detokenize{10.1088/0004-637X/774/1/73}}}.

\bibitem[{Kennicutt}(1984)]{Kennicutt84}
{Kennicutt}, Jr., R.C.
\newblock {Structural properties of giant H II regions in nearby galaxies}.
\newblock {\em \apj} {\bf 1984}, {\em 287},~116--130.
\newblock
doi:{\changeurlcolor{black}\href{https://doi.org/10.1086/162669}{\detokenize{10.1086/162669}}}.


\bibitem[{Parker}(1993)]{Parker93a}
{Parker}, J.W.
\newblock {The OB associations of 30 Doradus in the Large Magellanic Cloud. I -
  Stellar observations and data reductions}.
\newblock {\em \aj} {\bf 1993}, {\em 106},~560--577.
\newblock
  doi:{\changeurlcolor{black}\href{https://doi.org/10.1086/116661}{\detokenize{10.1086/116661}}}.

\bibitem[{Parker} and {Garmany}(1993)]{Parker93b}
{Parker}, J.W.; {Garmany}, C.D.
\newblock {The OB associations of 30 Doradus in the Large Magellanic Cloud. II
  - Stellar content and initial mass function}.
\newblock {\em \aj} {\bf 1993}, {\em 106},~1471--1483.
\newblock
  doi:{\changeurlcolor{black}\href{https://doi.org/10.1086/116740}{\detokenize{10.1086/116740}}}.


\bibitem[{Massey} and {Hunter}(1998)]{Massey98}
{Massey}, P.; {Hunter}, D.A.
\newblock {Star Formation in R136: A Cluster of O3 Stars Revealed by Hubble
  Space Telescope Spectroscopy}.
\newblock {\em \apj} {\bf 1998}, {\em 493},~180--194.
\newblock
  doi:{\changeurlcolor{black}\href{https://doi.org/10.1086/305126}{\detokenize{10.1086/305126}}}.

 
\bibitem[{Bosch} \em{et~al.}(1999){Bosch}, {Terlevich}, {Melnick}, and
  {Selman}]{Bosch99}
{Bosch}, G.; {Terlevich}, R.; {Melnick}, J.; {Selman}, F.
\newblock {The ionising cluster of 30 Doradus. II. Spectral classification for
  175 stars}.
\newblock {\em \aaps} {\bf 1999}, {\em 137},~21--41.
\newblock
  doi:{\changeurlcolor{black}\href{https://doi.org/10.1051/aas:1999480}{\detokenize{10.1051/aas:1999480}}}.



\bibitem[{Selman} \em{et~al.}(1999){Selman}, {Melnick}, {Bosch}, and
  {Terlevich}]{Selman99}
{Selman}, F.; {Melnick}, J.; {Bosch}, G.; {Terlevich}, R.
\newblock {The ionizing cluster of 30 Doradus. III. Star-formation history and
  initial mass function}.
\newblock {\em \aap} {\bf 1999}, {\em 347},~532--549.



\bibitem[{Breysacher} \em{et~al.}(1999){Breysacher}, {Azzopardi}, and {Testor}]{Breysacher99}
{Breysacher}, J.; {Azzopardi}, M.; {Testor}, G.
\newblock{The fourth catalogue of Population I Wolf-Rayet stars in the Large Magellanic Cloud}
\newblock {\em \aaps} {\bf 1999} {\em 137},~117-145

  
\bibitem[{Walborn} \em{et~al.}(2017){Walborn}, {Gamen}, {Morrell}, {Barb{\'a}},
  {Fern{\'a}ndez Laj{\'u}s}, and {Angeloni}]{Walborn17}
{Walborn}, N.R.; {Gamen}, R.C.; {Morrell}, N.I.; {Barb{\'a}}, R.H.;
  {Fern{\'a}ndez Laj{\'u}s}, E.; {Angeloni}, R.
\newblock {Active Luminous Blue Variables in the Large Magellanic Cloud}.
\newblock {\em \aj} {\bf 2017}, {\em 154},~15.
\newblock
  doi:{\changeurlcolor{black}\href{https://doi.org/10.3847/1538-3881/aa6195}{\detokenize{10.3847/1538-3881/aa6195}}}.

\bibitem[{Lennon} \em{et~al.}(2018){Lennon}, {Evans}, {van der Marel},
  {Anderson}, {Platais}, {Herrero}, {de Mink}, {Sana}, {Sabbi}, {Bedin},
  {Crowther}, {Langer}, {Ramos Lerate}, {del Pino}, {Renzo},
  {Sim{\'o}n-D{\'{\i}}az}, and {Schneider}]{Lennon18}
{Lennon}, D.J.; {Evans}, C.J.; {van der Marel}, R.P.; {Anderson}, J.;
  {Platais}, I.; {Herrero}, A.; {de Mink}, S.E.; {Sana}, H.; {Sabbi}, E.;
  {Bedin}, L.R. et al.
\newblock {Gaia DR2 reveals a very massive runaway star ejected from R136}.
\newblock {\em \aap} {\bf 2018}, {\em 619},~A78,
  \href{http://xxx.lanl.gov/abs/1805.08277}{{\normalfont
  [arXiv:astro-ph.SR/1805.08277]}}.
\newblock
  doi:{\changeurlcolor{black}\href{https://doi.org/10.1051/0004-6361/201833465}{\detokenize{10.1051/0004-6361/201833465}}}.

\bibitem[{Dufton} \em{et~al.}(2011){Dufton}, {Dunstall}, {Evans}, {Brott},
  {Cantiello}, {de Koter}, {de Mink}, {Fraser}, {H{\'e}nault-Brunet},
  {Howarth}, {Langer}, {Lennon}, {Markova}, {Sana}, and {Taylor}]{Dufton11}
{Dufton}, P.L.; {Dunstall}, P.R.; {Evans}, C.J.; {Brott}, I.; {Cantiello}, M.;
  {de Koter}, A.; {de Mink}, S.E.; {Fraser}, M.; {H{\'e}nault-Brunet}, V.;
  {Howarth}, I.D. et al.
\newblock {The VLT-FLAMES Tarantula Survey: The Fastest Rotating O-type Star
  and Shortest Period LMC Pulsar: Remnants of a Supernova Disrupted Binary?}
\newblock {\em \apjl} {\bf 2011}, {\em 743},~L22,
  \href{http://xxx.lanl.gov/abs/1111.0157}{{\normalfont
  [arXiv:astro-ph.SR/1111.0157]}}.
\newblock
  doi:{\changeurlcolor{black}\href{https://doi.org/10.1088/2041-8205/743/1/L22}{\detokenize{10.1088/2041-8205/743/1/L22}}}.

\bibitem[{Almeida} \em{et~al.}(2015){Almeida}, {Sana}, {de Mink}, {Tramper},
  {Soszy{\'n}ski}, {Langer}, {Barb{\'a}}, {Cantiello}, {Damineli}, {de Koter},
  {Garcia}, {Gr{\"a}fener}, {Herrero}, {Howarth}, {Ma{\'\i}z Apell{\'a}niz},
  {Norman}, {Ram{\'\i}rez-Agudelo}, and {Vink}]{Almeida15}
{Almeida}, L.A.; {Sana}, H.; {de Mink}, S.E.; {Tramper}, F.; {Soszy{\'n}ski},
  I.; {Langer}, N.; {Barb{\'a}}, R.H.; {Cantiello}, M.; {Damineli}, A.; {de
  Koter}, A. et al.
\newblock {Discovery of the Massive Overcontact Binary VFTS352: Evidence for
  Enhanced Internal Mixing}.
\newblock {\em \apj} {\bf 2015}, {\em 812},~102,
  \href{http://xxx.lanl.gov/abs/1509.08940}{{\normalfont
  [arXiv:astro-ph.SR/1509.08940]}}.
\newblock
  doi:{\changeurlcolor{black}\href{https://doi.org/10.1088/0004-637X/812/2/102}{\detokenize{10.1088/0004-637X/812/2/102}}}.

\bibitem[{Chen} \em{et~al.}(2006){Chen}, {Wang}, {Gotthelf}, {Jiang}, {Chu},
  and {Gruendl}]{Chen06}
{Chen}, Y.; {Wang}, Q.D.; {Gotthelf}, E.V.; {Jiang}, B.; {Chu}, Y.H.;
  {Gruendl}, R.
\newblock {Chandra ACIS Spectroscopy of N157B: A Young Composite Supernova
  Remnant in a Superbubble}.
\newblock {\em \apj} {\bf 2006}, {\em 651},~237--249,
  \href{http://xxx.lanl.gov/abs/astro-ph/0603123}{{\normalfont
  [astro-ph/0603123]}}.
\newblock
  doi:{\changeurlcolor{black}\href{https://doi.org/10.1086/507017}{\detokenize{10.1086/507017}}}.

\bibitem[{Savage} \em{et~al.}(1983){Savage}, {Fitzpatrick}, {Cassinelli}, and
  {Ebbets}]{Savage83}
{Savage}, B.D.; {Fitzpatrick}, E.L.; {Cassinelli}, J.P.; {Ebbets}, D.C.
\newblock {The nature of R136a, the superluminous central object of the 30
  Doradus nebula}.
\newblock {\em \apj} {\bf 1983}, {\em 273},~597--623.
\newblock
  doi:{\changeurlcolor{black}\href{https://doi.org/10.1086/161395}{\detokenize{10.1086/161395}}}.

\bibitem[{Weigelt} and {Baier}(1985)]{Weigelt85}
{Weigelt}, G.; {Baier}, G.
\newblock {R136a in the 30 Doradus nebula resolved by holographic speckle
  interferometry}.
\newblock {\em \aap} {\bf 1985}, {\em 150},~L18--L20.

\bibitem[{Hunter} \em{et~al.}(1995){Hunter}, {Shaya}, {Holtzman}, {Light},
  {O'Neil}, and {Lynds}]{Hunter95}
{Hunter}, D.A.; {Shaya}, E.J.; {Holtzman}, J.A.; {Light}, R.M.; {O'Neil}, Jr.,
  E.J.; {Lynds}, R.
\newblock {The Intermediate Stellar Mass Population in R136 Determined from
  Hubble Space Telescope Planetary Camera 2 Images}.
\newblock {\em \apj} {\bf 1995}, {\em 448},~179.
\newblock
  doi:{\changeurlcolor{black}\href{https://doi.org/10.1086/175950}{\detokenize{10.1086/175950}}}.

\bibitem[{de Koter} \em{et~al.}(1997){de Koter}, {Heap}, and
  {Hubeny}]{deKoter97}
{de Koter}, A.; {Heap}, S.R.; {Hubeny}, I.
\newblock {On the Evolutionary Phase and Mass Loss of the Wolf-Rayet--like
  Stars in R136a}.
\newblock {\em \apj} {\bf 1997}, {\em 477},~792.
\newblock
doi:{\changeurlcolor{black}\href{https://doi.org/10.1086/303736}{\detokenize{10.1086/303736}}}.

\bibitem[{Khorrami} \em{et~al.}(2017){Khorrami}, {Vakili}, {Lanz}, {Langlois},
  {Lagadec}, {Meyer}, {Robbe-Dubois}, {Abe}, {Avenhaus}, {Beuzit}, {Gratton},
  {Mouillet}, {Orign{\'e}}, {Petit}, and {Ramos}]{Khorrami17}
{Khorrami}, Z.; {Vakili}, F.; {Lanz}, T.; {Langlois}, M.; {Lagadec}, E.;
  {Meyer}, M.R.; {Robbe-Dubois}, S.; {Abe}, L.; {Avenhaus}, H.; {Beuzit}, J.L. et al.
\newblock {Uncrowding R 136 from VLT/SPHERE extreme adaptive optics}.
\newblock {\em \aap} {\bf 2017}, {\em 602},~A56,
  \href{http://xxx.lanl.gov/abs/1703.02876}{{\normalfont
  [arXiv:astro-ph.SR/1703.02876]}}.
\newblock
  doi:{\changeurlcolor{black}\href{https://doi.org/10.1051/0004-6361/201629279}{\detokenize{10.1051/0004-6361/201629279}}}.


\bibitem[{Tehrani} \em{et~al.}(2019){Tehrani}, {Crowther}, {Bestenlehner},
  {Littlefair}, {Pollock}, {Parker}, and {Schnurr}]{Tehrani19}
{Tehrani}, K.A.; {Crowther}, P.A.; {Bestenlehner}, J.M.; {Littlefair}, S.P.;
  {Pollock}, A.M.T.; {Parker}, R.J.; {Schnurr}, O.
\newblock {Weighing Melnick 34: the most massive binary system known}.
\newblock {\em \mnras} {\bf 2019}, {\em 484},~2692--2710,
  \href{http://xxx.lanl.gov/abs/1901.04769}{{\normalfont
  [arXiv:astro-ph.SR/1901.04769]}}.
\newblock
  doi:{\changeurlcolor{black}\href{https://doi.org/10.1093/mnras/stz147}{\detokenize{10.1093/mnras/stz147}}}.

\bibitem[{Pollock} \em{et~al.}(2018){Pollock}, {Crowther}, {Tehrani}, {Broos},
  and {Townsley}]{Pollock18}
{Pollock}, A.M.T.; {Crowther}, P.A.; {Tehrani}, K.; {Broos}, P.S.; {Townsley},
  L.K.
\newblock {The 155-day X-ray cycle of the very massive Wolf-Rayet star Melnick
  34 in the Large Magellanic Cloud}.
\newblock {\em \mnras} {\bf 2018}, {\em 474},~3228--3236,
  \href{http://xxx.lanl.gov/abs/1803.00822}{{\normalfont
  [arXiv:astro-ph.SR/1803.00822]}}.
\newblock
  doi:{\changeurlcolor{black}\href{https://doi.org/10.1093/mnras/stx2879}{\detokenize{10.1093/mnras/stx2879}}}.

\bibitem[{Smith}(2006)]{Smith06}
{Smith}, N.
\newblock {A census of the Carina Nebula - I. Cumulative energy input from
  massive stars}.
\newblock {\em \mnras} {\bf 2006}, {\em 367},~763--772,
  \href{http://xxx.lanl.gov/abs/astro-ph/0601060}{{\normalfont
  [astro-ph/0601060]}}.
\newblock
  doi:{\changeurlcolor{black}\href{https://doi.org/10.1111/j.1365-2966.2006.10007.x}{\detokenize{10.1111/j.1365-2966.2006.10007.x}}}.

\bibitem[{Schneider} \em{et~al.}(2018){Schneider}, {Sana}, {Evans},
  {Bestenlehner}, {Castro}, {Fossati}, {Gr{\"a}fener}, {Langer},
  {Ram{\'\i}rez-Agudelo}, {Sab{\'\i}n-Sanjuli{\'a}n}, {Sim{\'o}n-D{\'\i}az},
  {Tramper}, {Crowther}, {de Koter}, {de Mink}, {Dufton}, {Garcia}, {Gieles},
  {H{\'e}nault-Brunet}, {Herrero}, {Izzard}, {Kalari}, {Lennon}, {Ma{\'\i}z
  Apell{\'a}niz}, {Markova}, {Najarro}, {Podsiadlowski}, {Puls}, {Taylor}, {van
  Loon}, {Vink}, and {Norman}]{Schneider18a}
{Schneider}, F.R.N.; {Sana}, H.; {Evans}, C.J.; {Bestenlehner}, J.M.; {Castro},
  N.; {Fossati}, L.; {Gr{\"a}fener}, G.; {Langer}, N.; {Ram{\'\i}rez-Agudelo},
  O.H.; {Sab{\'\i}n-Sanjuli{\'a}n}, C. et al.
\newblock {An excess of massive stars in the local 30 Doradus starburst}.
\newblock {\em Science} {\bf 2018}, {\em 359},~69--71,
  \href{http://xxx.lanl.gov/abs/1801.03107}{{\normalfont
  [arXiv:astro-ph.SR/1801.03107]}}.
\newblock
  doi:{\changeurlcolor{black}\href{https://doi.org/10.1126/science.aan0106}{\detokenize{10.1126/science.aan0106}}}.

\bibitem[{Puls} \em{et~al.}(2005){Puls}, {Urbaneja}, {Venero}, {Repolust},
  {Springmann}, {Jokuthy}, and {Mokiem}]{Puls05}
{Puls}, J.; {Urbaneja}, M.A.; {Venero}, R.; {Repolust}, T.; {Springmann}, U.;
  {Jokuthy}, A.; {Mokiem}, M.R.
\newblock {Atmospheric NLTE-models for the spectroscopic analysis of blue stars
  with winds. II. Line-blanketed models}.
\newblock {\em \aap} {\bf 2005}, {\em 435},~669--698,
  \href{http://xxx.lanl.gov/abs/astro-ph/0411398}{{\normalfont
  [astro-ph/0411398]}}.
\newblock
  doi:{\changeurlcolor{black}\href{https://doi.org/10.1051/0004-6361:20042365}{\detokenize{10.1051/0004-6361:20042365}}}.

\bibitem[{Hillier} and {Miller}(1998)]{Hillier98}
{Hillier}, D.J.; {Miller}, D.L.
\newblock {The Treatment of Non-LTE Line Blanketing in Spherically Expanding
  Outflows}.
\newblock {\em \apj} {\bf 1998}, {\em 496},~407--427.
\newblock
doi:{\changeurlcolor{black}\href{https://doi.org/10.1086/305350}{\detokenize{10.1086/305350}}}.


\bibitem[{Gr{\"a}fener} \em{et~al.}(2002){Gr{\"a}fener}, {Koesterke}, and
  {Hamann}]{Grafener02}
{Gr{\"a}fener}, G.; {Koesterke}, L.; {Hamann}, W.R.
\newblock {Line-blanketed model atmospheres for WR stars}.
\newblock {\em \aap} {\bf 2002}, {\em 387},~244--257.
\newblock
  doi:{\changeurlcolor{black}\href{https://doi.org/10.1051/0004-6361:20020269}{\detokenize{10.1051/0004-6361:20020269}}}.



\bibitem[{Lanz} and {Hubeny}(2007)]{Lanz07}
{Lanz}, T.; {Hubeny}, I.
\newblock {A Grid of NLTE Line-blanketed Model Atmospheres of Early B-Type
  Stars}.
\newblock {\em \apjs} {\bf 2007}, {\em 169},~83--104,
  \href{http://xxx.lanl.gov/abs/astro-ph/0611891}{{\normalfont
  [astro-ph/0611891]}}.
\newblock
  doi:{\changeurlcolor{black}\href{https://doi.org/10.1086/511270}{\detokenize{10.1086/511270}}}.

\bibitem[{Gustafsson} \em{et~al.}(2008){Gustafsson}, {Edvardsson}, {Eriksson},
  {J{\o}rgensen}, {Nordlund}, and {Plez}]{Gustafsson08}
{Gustafsson}, B.; {Edvardsson}, B.; {Eriksson}, K.; {J{\o}rgensen}, U.G.;
  {Nordlund}, {\AA}.; {Plez}, B.
\newblock {A grid of MARCS model atmospheres for late-type stars. I. Methods
  and general properties}.
\newblock {\em \aap} {\bf 2008}, {\em 486},~951--970,
  \href{http://xxx.lanl.gov/abs/0805.0554}{{\normalfont [0805.0554]}}.
\newblock
  doi:{\changeurlcolor{black}\href{https://doi.org/10.1051/0004-6361:200809724}{\detokenize{10.1051/0004-6361:200809724}}}.

\bibitem[{Langer} and {Kudritzki}(2014)]{Langer14}
{Langer}, N.; {Kudritzki}, R.P.
\newblock {The spectroscopic Hertzsprung-Russell diagram}.
\newblock {\em \aap} {\bf 2014}, {\em 564},~A52,
  \href{http://xxx.lanl.gov/abs/1403.2212}{{\normalfont
  [arXiv:astro-ph.SR/1403.2212]}}.
\newblock
  doi:{\changeurlcolor{black}\href{https://doi.org/10.1051/0004-6361/201423374}{\detokenize{10.1051/0004-6361/201423374}}}.

\bibitem[{Bestenlehner} \em{et~al.}(2014){Bestenlehner}, {Gr{\"a}fener},
  {Vink}, {Najarro}, {de Koter}, {Sana}, {Evans}, {Crowther},
  {H{\'e}nault-Brunet}, {Herrero}, {Langer}, {Schneider},
  {Sim{\'o}n-D{\'{\i}}az}, {Taylor}, and {Walborn}]{Bestenlehner14}
{Bestenlehner}, J.M.; {Gr{\"a}fener}, G.; {Vink}, J.S.; {Najarro}, F.; {de
  Koter}, A.; {Sana}, H.; {Evans}, C.J.; {Crowther}, P.A.;
  {H{\'e}nault-Brunet}, V.; {Herrero}, A. et al.
\newblock {The VLT-FLAMES Tarantula Survey. XVII. Physical and wind properties
  of massive stars at the top of the main sequence}.
\newblock {\em \aap} {\bf 2014}, {\em 570},~A38,
  \href{http://xxx.lanl.gov/abs/1407.1837}{{\normalfont
  [arXiv:astro-ph.SR/1407.1837]}}.
\newblock
  doi:{\changeurlcolor{black}\href{https://doi.org/10.1051/0004-6361/201423643}{\detokenize{10.1051/0004-6361/201423643}}}.

\bibitem[{McEvoy} \em{et~al.}(2015){McEvoy}, {Dufton}, {Evans}, {Kalari},
  {Markova}, {Sim{\'o}n-D{\'{\i}}az}, {Vink}, {Walborn}, {Crowther}, {de
  Koter}, {de Mink}, {Dunstall}, {H{\'e}nault-Brunet}, {Herrero}, {Langer},
  {Lennon}, {Ma{\'{\i}}z Apell{\'a}niz}, {Najarro}, {Puls}, {Sana},
  {Schneider}, and {Taylor}]{McEvoy15}
{McEvoy}, C.M.; {Dufton}, P.L.; {Evans}, C.J.; {Kalari}, V.M.; {Markova}, N.;
  {Sim{\'o}n-D{\'{\i}}az}, S.; {Vink}, J.S.; {Walborn}, N.R.; {Crowther}, P.A.;
  {de Koter}, A. et al.
\newblock {The VLT-FLAMES Tarantula Survey. XIX. B-type supergiants:
  Atmospheric parameters and nitrogen abundances to investigate the role of
  binarity and the width of the main sequence}.
\newblock {\em \aap} {\bf 2015}, {\em 575},~A70,
  \href{http://xxx.lanl.gov/abs/1412.2705}{{\normalfont
  [arXiv:astro-ph.SR/1412.2705]}}.
\newblock
  doi:{\changeurlcolor{black}\href{https://doi.org/10.1051/0004-6361/201425202}{\detokenize{10.1051/0004-6361/201425202}}}.

\bibitem[{Sab{\'{\i}}n-Sanjuli{\'a}n}
  \em{et~al.}(2017){Sab{\'{\i}}n-Sanjuli{\'a}n}, {Sim{\'o}n-D{\'{\i}}az},
  {Herrero}, {Puls}, {Schneider}, {Evans}, {Garcia}, {Najarro}, {Brott},
  {Castro}, {Crowther}, {de Koter}, {de Mink}, {Gr{\"a}fener}, {Grin},
  {Holgado}, {Langer}, {Lennon}, {Ma{\'{\i}}z Apell{\'a}niz},
  {Ram{\'{\i}}rez-Agudelo}, {Sana}, {Taylor}, {Vink}, and
  {Walborn}]{SabinSanjulian17}
{Sab{\'{\i}}n-Sanjuli{\'a}n}, C.; {Sim{\'o}n-D{\'{\i}}az}, S.; {Herrero}, A.;
  {Puls}, J.; {Schneider}, F.R.N.; {Evans}, C.J.; {Garcia}, M.; {Najarro}, F.;
  {Brott}, I.; {Castro}, N. et al.
\newblock {The VLT-FLAMES Tarantula Survey. XXVI. Properties of the O-dwarf
  population in 30 Doradus}.
\newblock {\em \aap} {\bf 2017}, {\em 601},~A79,
  \href{http://xxx.lanl.gov/abs/1702.04773}{{\normalfont
  [arXiv:astro-ph.SR/1702.04773]}}.
\newblock
  doi:{\changeurlcolor{black}\href{https://doi.org/10.1051/0004-6361/201629210}{\detokenize{10.1051/0004-6361/201629210}}}.

\bibitem[{Ram{\'{\i}}rez-Agudelo} \em{et~al.}(2017){Ram{\'{\i}}rez-Agudelo},
  {Sana}, {de Koter}, {Tramper}, {Grin}, {Schneider}, {Langer}, {Puls},
  {Markova}, {Bestenlehner}, {Castro}, {Crowther}, {Evans}, {Garc{\'{\i}}a},
  {Gr{\"a}fener}, {Herrero}, {van Kempen}, {Lennon}, {Ma{\'{\i}}z
  Apell{\'a}niz}, {Najarro}, {Sab{\'{\i}}n-Sanjuli{\'a}n},
  {Sim{\'o}n-D{\'{\i}}az}, {Taylor}, and {Vink}]{RamirezAgudelo17}
{Ram{\'{\i}}rez-Agudelo}, O.H.; {Sana}, H.; {de Koter}, A.; {Tramper}, F.;
  {Grin}, N.J.; {Schneider}, F.R.N.; {Langer}, N.; {Puls}, J.; {Markova}, N.;
  {Bestenlehner}, J.M. et al.
\newblock {The VLT-FLAMES Tarantula Survey . XXIV. Stellar properties of the
  O-type giants and supergiants in 30 Doradus}.
\newblock {\em \aap} {\bf 2017}, {\em 600},~A81,
  \href{http://xxx.lanl.gov/abs/1701.04758}{{\normalfont
  [arXiv:astro-ph.SR/1701.04758]}}.
\newblock
  doi:{\changeurlcolor{black}\href{https://doi.org/10.1051/0004-6361/201628914}{\detokenize{10.1051/0004-6361/201628914}}}.

\bibitem[{Dufton} \em{et~al.}(2018){Dufton}, {Thompson}, {Crowther}, {Evans},
  {Schneider}, {de Koter}, {de Mink}, {Garland}, {Langer}, {Lennon}, {McEvoy},
  {Ram{\'{\i}}rez-Agudelo}, {Sana}, {S{\'{\i}}mon D{\'{\i}}az}, {Taylor}, and
  {Vink}]{Dufton18}
{Dufton}, P.L.; {Thompson}, A.; {Crowther}, P.A.; {Evans}, C.J.; {Schneider},
  F.R.N.; {de Koter}, A.; {de Mink}, S.E.; {Garland}, R.; {Langer}, N.;
  {Lennon}, D.J. et al.
\newblock {The VLT-FLAMES Tarantula Survey. XXVIII. Nitrogen abundances for
  apparently single dwarf and giant B-type stars with small projected
  rotational velocities}.
\newblock {\em \aap} {\bf 2018}, {\em 615},~A101,
  \href{http://xxx.lanl.gov/abs/1804.02025}{{\normalfont
  [arXiv:astro-ph.SR/1804.02025]}}.
\newblock
  doi:{\changeurlcolor{black}\href{https://doi.org/10.1051/0004-6361/201732440}{\detokenize{10.1051/0004-6361/201732440}}}.

\bibitem[{Bestenlehner} \em{et~al.}("in prep."){Bestenlehner}, {Crowther},
  {Caballero-Nieves}, {Sim\'{o}n-D\'{i}az}, and {Schneider}]{Bestenlehner19b}
{Bestenlehner}, J.M.; {Crowther}, P.; {Caballero-Nieves}, S.M.;
  {Sim\'{o}n-D\'{i}az}, S.; {Schneider}, F.
\newblock {The R136 star cluster dissected with Hubble Space Telescope.III
  Physical properties of the most massive stars in R136}.
\newblock {\em \mnras} {\bf in prep.}.


\bibitem[{Brott} \em{et~al.}(2011){Brott}, {de Mink}, {Cantiello}, {Langer},
  {de Koter}, {Evans}, {Hunter}, {Trundle}, and {Vink}]{Brott11}
{Brott}, I.; {de Mink}, S.E.; {Cantiello}, M.; {Langer}, N.; {de Koter}, A.;
  {Evans}, C.J.; {Hunter}, I.; {Trundle}, C.; {Vink}, J.S.
\newblock {Rotating massive main-sequence stars. I. Grids of evolutionary
  models and isochrones}.
\newblock {\em \aap} {\bf 2011}, {\em 530},~A115,
  \href{http://xxx.lanl.gov/abs/1102.0530}{{\normalfont
  [arXiv:astro-ph.SR/1102.0530]}}.
\newblock
  doi:{\changeurlcolor{black}\href{https://doi.org/10.1051/0004-6361/201016113}{\detokenize{10.1051/0004-6361/201016113}}}.

\bibitem[{K{\"o}hler} \em{et~al.}(2015){K{\"o}hler}, {Langer}, {de Koter}, {de
  Mink}, {Crowther}, {Evans}, {Gr{\"a}fener}, {Sana}, {Sanyal}, {Schneider},
  and {Vink}]{Kohler15}
{K{\"o}hler}, K.; {Langer}, N.; {de Koter}, A.; {de Mink}, S.E.; {Crowther},
  P.A.; {Evans}, C.J.; {Gr{\"a}fener}, G.; {Sana}, H.; {Sanyal}, D.;
  {Schneider}, F.R.N.; {Vink}, J.S.
\newblock {The evolution of rotating very massive stars with LMC composition}.
\newblock {\em \aap} {\bf 2015}, {\em 573},~A71,
  \href{http://xxx.lanl.gov/abs/1501.03794}{{\normalfont
  [arXiv:astro-ph.SR/1501.03794]}}.
\newblock
  doi:{\changeurlcolor{black}\href{https://doi.org/10.1051/0004-6361/201424356}{\detokenize{10.1051/0004-6361/201424356}}}.

\bibitem[{Gordon} \em{et~al.}(2016){Gordon}, {Humphreys}, and
  {Jones}]{Gordon16}
{Gordon}, M.S.; {Humphreys}, R.M.; {Jones}, T.J.
\newblock {Luminous and Variable Stars in M31 and M33. III. The Yellow and Red
  Supergiants and Post-red Supergiant Evolution}.
\newblock {\em \apj} {\bf 2016}, {\em 825},~50,
  \href{http://xxx.lanl.gov/abs/1603.08003}{{\normalfont
  [arXiv:astro-ph.SR/1603.08003]}}.
\newblock
  doi:{\changeurlcolor{black}\href{https://doi.org/10.3847/0004-637X/825/1/50}{\detokenize{10.3847/0004-637X/825/1/50}}}.

\bibitem[{Davies} \em{et~al.}(2018){Davies}, {Crowther}, and
  {Beasor}]{Davies18}
{Davies}, B.; {Crowther}, P.A.; {Beasor}, E.R.
\newblock {The luminosities of cool supergiants in the Magellanic Clouds, and
  the Humphreys-Davidson limit revisited}.
\newblock {\em \mnras} {\bf 2018}, {\em 478},~3138--3148,
  \href{http://xxx.lanl.gov/abs/1804.06417}{{\normalfont
  [arXiv:astro-ph.SR/1804.06417]}}.
\newblock
  doi:{\changeurlcolor{black}\href{https://doi.org/10.1093/mnras/sty1302}{\detokenize{10.1093/mnras/sty1302}}}.

\bibitem[{Schneider} \em{et~al.}(2014){Schneider}, {Langer}, {de Koter},
  {Brott}, {Izzard}, and {Lau}]{Schneider14}
{Schneider}, F.R.N.; {Langer}, N.; {de Koter}, A.; {Brott}, I.; {Izzard}, R.G.;
  {Lau}, H.H.B.
\newblock {Bonnsai: a Bayesian tool for comparing stars with stellar evolution
  models}.
\newblock {\em \aap} {\bf 2014}, {\em 570},~A66,
  \href{http://xxx.lanl.gov/abs/1408.3409}{{\normalfont
  [arXiv:astro-ph.SR/1408.3409]}}.
\newblock
  doi:{\changeurlcolor{black}\href{https://doi.org/10.1051/0004-6361/201424286}{\detokenize{10.1051/0004-6361/201424286}}}.

\bibitem[{Sana} \em{et~al.}(2012){Sana}, {de Mink}, {de Koter}, {Langer},
  {Evans}, {Gieles}, {Gosset}, {Izzard}, {Le Bouquin}, and {Schneider}]{Sana12}
{Sana}, H.; {de Mink}, S.E.; {de Koter}, A.; {Langer}, N.; {Evans}, C.J.;
  {Gieles}, M.; {Gosset}, E.; {Izzard}, R.G.; {Le Bouquin}, J.B.; {Schneider},
  F.R.N.
\newblock {Binary Interaction Dominates the Evolution of Massive Stars}.
\newblock {\em Science} {\bf 2012}, {\em 337},~444,
  \href{http://xxx.lanl.gov/abs/1207.6397}{{\normalfont
  [arXiv:astro-ph.SR/1207.6397]}}.
\newblock
  doi:{\changeurlcolor{black}\href{https://doi.org/10.1126/science.1223344}{\detokenize{10.1126/science.1223344}}}.

\bibitem[{Sana} \em{et~al.}(2013){Sana}, {de Koter}, {de Mink}, {Dunstall},
  {Evans}, {H{\'e}nault-Brunet}, {Ma{\'{\i}}z Apell{\'a}niz},
  {Ram{\'{\i}}rez-Agudelo}, {Taylor}, {Walborn}, {Clark}, {Crowther},
  {Herrero}, {Gieles}, {Langer}, {Lennon}, and {Vink}]{Sana13}
{Sana}, H.; {de Koter}, A.; {de Mink}, S.E.; {Dunstall}, P.R.; {Evans}, C.J.;
  {H{\'e}nault-Brunet}, V.; {Ma{\'{\i}}z Apell{\'a}niz}, J.;
  {Ram{\'{\i}}rez-Agudelo}, O.H.; {Taylor}, W.D.; {Walborn}, N.R. et al.
\newblock {The VLT-FLAMES Tarantula Survey. VIII. Multiplicity properties of
  the O-type star population}.
\newblock {\em \aap} {\bf 2013}, {\em 550},~A107,
  \href{http://xxx.lanl.gov/abs/1209.4638}{{\normalfont
  [arXiv:astro-ph.SR/1209.4638]}}.
\newblock
  doi:{\changeurlcolor{black}\href{https://doi.org/10.1051/0004-6361/201219621}{\detokenize{10.1051/0004-6361/201219621}}}.


\bibitem[{Hainich} \em{et~al.}(2014){Hainich}, {R{\"u}hling}, {Todt},
  {Oskinova}, {Liermann}, {Gr{\"a}fener}, {Foellmi}, {Schnurr}, and
  {Hamann}]{Hainich14}
{Hainich}, R.; {R{\"u}hling}, U.; {Todt}, H.; {Oskinova}, L.M.; {Liermann}, A.;
  {Gr{\"a}fener}, G.; {Foellmi}, C.; {Schnurr}, O.; {Hamann}, W.R.
\newblock {The Wolf-Rayet stars in the Large Magellanic Cloud. A comprehensive
  analysis of the WN class}.
\newblock {\em \aap} {\bf 2014}, {\em 565},~A27,
  \href{http://xxx.lanl.gov/abs/1401.5474}{{\normalfont
  [arXiv:astro-ph.SR/1401.5474]}}.
\newblock
  doi:{\changeurlcolor{black}\href{https://doi.org/10.1051/0004-6361/201322696}{\detokenize{10.1051/0004-6361/201322696}}}.

  
\bibitem[{Garland} \em{et~al.}(2017){Garland}, {Dufton}, {Evans}, {Crowther},
  {Howarth}, {de Koter}, {de Mink}, {Grin}, {Langer}, {Lennon}, {McEvoy},
  {Sana}, {Schneider}, {S{\'{\i}}mon D{\'{\i}}az}, {Taylor}, {Thompson}, and
  {Vink}]{Garland17}
{Garland}, R.; {Dufton}, P.L.; {Evans}, C.J.; {Crowther}, P.A.; {Howarth},
  I.D.; {de Koter}, A.; {de Mink}, S.E.; {Grin}, N.J.; {Langer}, N.; {Lennon},
  D.J. et al.
\newblock {The VLT-FLAMES Tarantula Survey. XXVII. Physical parameters of
  B-type main-sequence binary systems in the Tarantula nebula}.
\newblock {\em \aap} {\bf 2017}, {\em 603},~A91,
  \href{http://xxx.lanl.gov/abs/1704.07131}{{\normalfont
  [arXiv:astro-ph.SR/1704.07131]}}.
\newblock
  doi:{\changeurlcolor{black}\href{https://doi.org/10.1051/0004-6361/201629982}{\detokenize{10.1051/0004-6361/201629982}}}.

\bibitem[{Mahy} \em{et~al.}("submitted"){Mahy}, {Sana}, {Abdul-Masih},
  {Almeida}, and {Langer}]{Mahy19}
{Mahy}, L.; {Sana}, H.; {Abdul-Masih}, M.; {Almeida}, L.A.; {Langer}, N.; {Shenar}, T.; {de 
Koter}, A.; {de Mink}, S.; {de Wit}, S.; {Grin}, N. et al.
\newblock {The Tarantula Massive Binary Monitoring: III. Atmosphere analysis of
  double-lined spectroscopic systems}.
\newblock {\em \aap} {\bf submitted}.

\bibitem[{Shenar} \em{et~al.}(2019){Shenar}, {Sablowski}, {Hainich}, {Todt},
  {Moffat}, {Oskinova}, {Ramachandran}, {Sana}, {Sander}, {Schnurr},
  {St-Louis}, {Vanbeveren}, {G{\"o}tberg}, and {Hamann}]{Shenar19}
{Shenar}, T.; {Sablowski}, D.P.; {Hainich}, R.; {Todt}, H.; {Moffat}, A.F.J.;
  {Oskinova}, L.M.; {Ramachandran}, V.; {Sana}, H.; {Sander}, A.A.C.;
  {Schnurr}, O. et al.
\newblock {The Wolf-Rayet binaries of the nitrogen sequence in the Large
  Magellanic Cloud. Spectroscopy, orbital analysis, formation, and evolution}.
\newblock {\em \aap} {\bf 2019}, {\em 627},~A151,
  \href{http://xxx.lanl.gov/abs/1905.09296}{{\normalfont
  [arXiv:astro-ph.SR/1905.09296]}}.
\newblock
  doi:{\changeurlcolor{black}\href{https://doi.org/10.1051/0004-6361/201935684}{\detokenize{10.1051/0004-6361/201935684}}}.

\bibitem[{Vanbeveren} \em{et~al.}(1998){Vanbeveren}, {De Loore}, and {Van
  Rensbergen}]{Vanbeveren98}
{Vanbeveren}, D.; {De Loore}, C.; {Van Rensbergen}, W.
\newblock {Massive stars}.
\newblock {\em \aapr} {\bf 1998}, {\em 9},~63--152.
\newblock
  doi:{\changeurlcolor{black}\href{https://doi.org/10.1007/s001590050015}{\detokenize{10.1007/s001590050015}}}.

\bibitem[{Dunstall} \em{et~al.}(2015){Dunstall}, {Dufton}, {Sana}, {Evans},
  {Howarth}, {Sim{\'o}n-D{\'{\i}}az}, {de Mink}, {Langer}, {Ma{\'{\i}}z
  Apell{\'a}niz}, and {Taylor}]{Dunstall15}
{Dunstall}, P.R.; {Dufton}, P.L.; {Sana}, H.; {Evans}, C.J.; {Howarth}, I.D.;
  {Sim{\'o}n-D{\'{\i}}az}, S.; {de Mink}, S.E.; {Langer}, N.; {Ma{\'{\i}}z
  Apell{\'a}niz}, J.; {Taylor}, W.D.
\newblock {The VLT-FLAMES Tarantula Survey. XXII. Multiplicity properties of
  the B-type stars}.
\newblock {\em \aap} {\bf 2015}, {\em 580},~A93,
  \href{http://xxx.lanl.gov/abs/1505.07121}{{\normalfont
  [arXiv:astro-ph.SR/1505.07121]}}.
\newblock
  doi:{\changeurlcolor{black}\href{https://doi.org/10.1051/0004-6361/201526192}{\detokenize{10.1051/0004-6361/201526192}}}.

\bibitem[{Almeida} \em{et~al.}(2017){Almeida}, {Sana}, {Taylor}, {Barb{\'a}},
  {Bonanos}, {Crowther}, {Damineli}, {de Koter}, {de Mink}, {Evans}, {Gieles},
  {Grin}, {H{\'e}nault-Brunet}, {Langer}, {Lennon}, {Lockwood}, {Ma{\'{\i}}z
  Apell{\'a}niz}, {Moffat}, {Neijssel}, {Norman}, {Ram{\'{\i}}rez-Agudelo},
  {Richardson}, {Schootemeijer}, {Shenar}, {Soszy{\'n}ski}, {Tramper}, and
  {Vink}]{Almeida17}
{Almeida}, L.A.; {Sana}, H.; {Taylor}, W.; {Barb{\'a}}, R.; {Bonanos}, A.Z.;
  {Crowther}, P.; {Damineli}, A.; {de Koter}, A.; {de Mink}, S.E.; {Evans},
  C.J. et al.
\newblock {The Tarantula Massive Binary Monitoring. I. Observational campaign
  and OB-type spectroscopic binaries}.
\newblock {\em \aap} {\bf 2017}, {\em 598},~A84,
  \href{http://xxx.lanl.gov/abs/1610.03500}{{\normalfont
  [arXiv:astro-ph.SR/1610.03500]}}.
\newblock
  doi:{\changeurlcolor{black}\href{https://doi.org/10.1051/0004-6361/201629844}{\detokenize{10.1051/0004-6361/201629844}}}.

\bibitem[{G{\"o}tberg} \em{et~al.}(2018){G{\"o}tberg}, {de Mink}, {Groh},
  {Kupfer}, {Crowther}, {Zapartas}, and {Renzo}]{Gotberg18}
{G{\"o}tberg}, Y.; {de Mink}, S.E.; {Groh}, J.H.; {Kupfer}, T.; {Crowther},
  P.A.; {Zapartas}, E.; {Renzo}, M.
\newblock {Spectral models for binary products: Unifying subdwarfs and
  Wolf-Rayet stars as a sequence of stripped-envelope stars}.
\newblock {\em \aap} {\bf 2018}, {\em 615},~A78,
  \href{http://xxx.lanl.gov/abs/1802.03018}{{\normalfont
  [arXiv:astro-ph.SR/1802.03018]}}.
\newblock
  doi:{\changeurlcolor{black}\href{https://doi.org/10.1051/0004-6361/201732274}{\detokenize{10.1051/0004-6361/201732274}}}.

\bibitem[{Ram{\'{\i}}rez-Agudelo} \em{et~al.}(2013){Ram{\'{\i}}rez-Agudelo},
  {Sim{\'o}n-D{\'{\i}}az}, {Sana}, {de Koter}, {Sab{\'{\i}}n-Sanjul{\'{\i}}an},
  {de Mink}, {Dufton}, {Gr{\"a}fener}, {Evans}, {Herrero}, {Langer}, {Lennon},
  {Ma{\'{\i}}z Apell{\'a}niz}, {Markova}, {Najarro}, {Puls}, {Taylor}, and
  {Vink}]{RamirezAgudelo13}
{Ram{\'{\i}}rez-Agudelo}, O.H.; {Sim{\'o}n-D{\'{\i}}az}, S.; {Sana}, H.; {de
  Koter}, A.; {Sab{\'{\i}}n-Sanjul{\'{\i}}an}, C.; {de Mink}, S.E.; {Dufton},
  P.L.; {Gr{\"a}fener}, G.; {Evans}, C.J.; {Herrero}, A.; {Langer}, N. et al.
\newblock {The VLT-FLAMES Tarantula Survey. XII. Rotational velocities of the
  single O-type stars}.
\newblock {\em \aap} {\bf 2013}, {\em 560},~A29,
  \href{http://xxx.lanl.gov/abs/1309.2929}{{\normalfont
  [arXiv:astro-ph.SR/1309.2929]}}.
\newblock
  doi:{\changeurlcolor{black}\href{https://doi.org/10.1051/0004-6361/201321986}{\detokenize{10.1051/0004-6361/201321986}}}.

\bibitem[{Dufton} \em{et~al.}(2013){Dufton}, {Langer}, {Dunstall}, {Evans},
  {Brott}, {de Mink}, {Howarth}, {Kennedy}, {McEvoy}, {Potter},
  {Ram{\'{\i}}rez-Agudelo}, {Sana}, {Sim{\'o}n-D{\'{\i}}az}, {Taylor}, and
  {Vink}]{Dufton13}
{Dufton}, P.L.; {Langer}, N.; {Dunstall}, P.R.; {Evans}, C.J.; {Brott}, I.; {de
  Mink}, S.E.; {Howarth}, I.D.; {Kennedy}, M.; {McEvoy}, C.; {Potter}, A.T. et al.
\newblock {The VLT-FLAMES Tarantula Survey. X. Evidence for a bimodal
  distribution of rotational velocities for the single early B-type stars}.
\newblock {\em \aap} {\bf 2013}, {\em 550},~A109,
  \href{http://xxx.lanl.gov/abs/1212.2424}{{\normalfont
  [arXiv:astro-ph.SR/1212.2424]}}.
\newblock
  doi:{\changeurlcolor{black}\href{https://doi.org/10.1051/0004-6361/201220273}{\detokenize{10.1051/0004-6361/201220273}}}.


\bibitem[{Sim{\'o}n-D{\'{\i}}az} and {Herrero}(2014)]{SimonDiaz14}
{Sim{\'o}n-D{\'{\i}}az}, S.; {Herrero}, A.
\newblock {The IACOB project. I. Rotational velocities in northern Galactic O-
  and early B-type stars revisited. The impact of other sources of
  line-broadening}.
\newblock {\em \aap} {\bf 2014}, {\em 562},~A135,
  \href{http://xxx.lanl.gov/abs/1311.3360}{{\normalfont
  [arXiv:astro-ph.SR/1311.3360]}}.
\newblock
  doi:{\changeurlcolor{black}\href{https://doi.org/10.1051/0004-6361/201322758}{\detokenize{10.1051/0004-6361/201322758}}}.

\bibitem[{Ram{\'{\i}}rez-Agudelo} \em{et~al.}(2015){Ram{\'{\i}}rez-Agudelo},
  {Sana}, {de Mink}, {H{\'e}nault-Brunet}, {de Koter}, {Langer}, {Tramper},
  {Gr{\"a}fener}, {Evans}, {Vink}, {Dufton}, and {Taylor}]{RamirezAgudelo15}
{Ram{\'{\i}}rez-Agudelo}, O.H.; {Sana}, H.; {de Mink}, S.E.;
  {H{\'e}nault-Brunet}, V.; {de Koter}, A.; {Langer}, N.; {Tramper}, F.;
  {Gr{\"a}fener}, G.; {Evans}, C.J.; {Vink}, J.S. et al.
\newblock {The VLT-FLAMES Tarantula Survey. XXI. Stellar spin rates of O-type
  spectroscopic binaries}.
\newblock {\em \aap} {\bf 2015}, {\em 580},~A92,
  \href{http://xxx.lanl.gov/abs/1507.02286}{{\normalfont
  [arXiv:astro-ph.SR/1507.02286]}}.
\newblock
  doi:{\changeurlcolor{black}\href{https://doi.org/10.1051/0004-6361/201425424}{\detokenize{10.1051/0004-6361/201425424}}}.

\bibitem[{Wolff} \em{et~al.}(2008){Wolff}, {Strom}, {Cunha}, {Daflon}, {Olsen},
  and {Dror}]{Wolff08}
{Wolff}, S.C.; {Strom}, S.E.; {Cunha}, K.; {Daflon}, S.; {Olsen}, K.; {Dror},
  D.
\newblock {Rotational Velocities for Early-Type Stars in the Young Large
  Magellanic Cloud Cluster R136: Further Study of the Relationship Between
  Rotation Speed and Density in Star-Forming Regions}.
\newblock {\em \aj} {\bf 2008}, {\em 136},~1049--1060.
\newblock
  doi:{\changeurlcolor{black}\href{https://doi.org/10.1088/0004-6256/136/3/1049}{\detokenize{10.1088/0004-6256/136/3/1049}}}.

\bibitem[{Platais} \em{et~al.}(2018){Platais}, {Lennon}, {van der Marel},
  {Bellini}, {Sabbi}, {Watkins}, {Sohn}, {Walborn}, {Bedin}, {Evans}, {de
  Mink}, {Sana}, {Herrero}, {Langer}, and {Crowther}]{Platais18}
{Platais}, I.; {Lennon}, D.J.; {van der Marel}, R.P.; {Bellini}, A.; {Sabbi},
  E.; {Watkins}, L.L.; {Sohn}, S.T.; {Walborn}, N.R.; {Bedin}, L.R.; {Evans},
  C.J.; {de Mink}, S.E. et al.
\newblock {HST Astrometry in the 30 Doradus Region. II. Runaway Stars from New
  Proper Motions in the Large Magellanic Cloud}.
\newblock {\em \aj} {\bf 2018}, {\em 156},~98,
  \href{http://xxx.lanl.gov/abs/1804.08678}{{\normalfont
  [arXiv:astro-ph.SR/1804.08678]}}.
\newblock
  doi:{\changeurlcolor{black}\href{https://doi.org/10.3847/1538-3881/aad280}{\detokenize{10.3847/1538-3881/aad280}}}.

\bibitem[{Evans} \em{et~al.}(2010){Evans}, {Walborn}, {Crowther},
  {H{\'e}nault-Brunet}, {Massa}, {Taylor}, {Howarth}, {Sana}, {Lennon}, and
  {van Loon}]{Evans10}
{Evans}, C.J.; {Walborn}, N.R.; {Crowther}, P.A.; {H{\'e}nault-Brunet}, V.;
  {Massa}, D.; {Taylor}, W.D.; {Howarth}, I.D.; {Sana}, H.; {Lennon}, D.J.;
  {van Loon}, J.T.
\newblock {A Massive Runaway Star from 30 Doradus}.
\newblock {\em \apjl} {\bf 2010}, {\em 715},~L74--L79,
  \href{http://xxx.lanl.gov/abs/1004.5402}{{\normalfont
  [arXiv:astro-ph.SR/1004.5402]}}.
\newblock
  doi:{\changeurlcolor{black}\href{https://doi.org/10.1088/2041-8205/715/2/L74}{\detokenize{10.1088/2041-8205/715/2/L74}}}.

\bibitem[{Renzo} \em{et~al.}(2019){Renzo}, {de Mink}, {Lennon}, {Platais}, {van
  der Marel}, {Laplace}, {Bestenlehner}, {Evans}, {H{\'e}nault-Brunet},
  {Justham}, {de Koter}, {Langer}, {Najarro}, {Schneider}, and {Vink}]{Renzo19}
{Renzo}, M.; {de Mink}, S.E.; {Lennon}, D.J.; {Platais}, I.; {van der Marel},
  R.P.; {Laplace}, E.; {Bestenlehner}, J.M.; {Evans}, C.J.;
  {H{\'e}nault-Brunet}, V.; {Justham}, S. et al.
\newblock {Space astrometry of the very massive 150 M$_{\odot}$ candidate
  runaway star VFTS682}.
\newblock {\em \mnras} {\bf 2019}, {\em 482},~L102--L106,
  \href{http://xxx.lanl.gov/abs/1810.05650}{{\normalfont
  [arXiv:astro-ph.SR/1810.05650]}}.
\newblock
doi:{\changeurlcolor{black}\href{https://doi.org/10.1093/mnrasl/sly194}{\detokenize{10.1093/mnrasl/sly194}}}.


\bibitem[{Pallavicini} \em{et~al.}(1981){Pallavicini}, {Golub}, {Rosner},
  {Vaiana}, {Ayres}, and {Linsky}]{Pallavicini81}
{Pallavicini}, R.; {Golub}, L.; {Rosner}, R.; {Vaiana}, G.S.; {Ayres}, T.;
  {Linsky}, J.L.
\newblock {Relations among stellar X-ray emission observed from Einstein,
  stellar rotation and bolometric luminosity}.
\newblock {\em \apj} {\bf 1981}, {\em 248},~279--290.
\newblock
  doi:{\changeurlcolor{black}\href{https://doi.org/10.1086/159152}{\detokenize{10.1086/159152}}}.


\bibitem[{Stevens} \em{et~al.}(1992){Stevens}, {Blondin}, and
  {Pollock}]{Stevens92}
{Stevens}, I.R.; {Blondin}, J.M.; {Pollock}, A.M.T.
\newblock {Colliding winds from early-type stars in binary systems}.
\newblock {\em \apj} {\bf 1992}, {\em 386},~265--287.
\newblock
  doi:{\changeurlcolor{black}\href{https://doi.org/10.1086/171013}{\detokenize{10.1086/171013}}}.


\bibitem[{Massey} \em{et~al.}(2002){Massey}, {Penny}, and {Vukovich}]{Massey02}
{Massey}, P.; {Penny}, L.R.; {Vukovich}, J.
\newblock {Orbits of Four Very Massive Binaries in the R136 Cluster}.
\newblock {\em \apj} {\bf 2002}, {\em 565},~982--993,
  \href{http://xxx.lanl.gov/abs/astro-ph/0110088}{{\normalfont
  [astro-ph/0110088]}}.
\newblock
  doi:{\changeurlcolor{black}\href{https://doi.org/10.1086/324783}{\detokenize{10.1086/324783}}}.

\bibitem[{Taylor} \em{et~al.}(2011){Taylor}, {Evans}, {Sana}, {Walborn}, {de
  Mink}, {Stroud}, {Alvarez-Candal}, {Barb{\'a}}, {Bestenlehner}, {Bonanos},
  {Brott}, {Crowther}, {de Koter}, {Friedrich}, {Gr{\"a}fener},
  {H{\'e}nault-Brunet}, {Herrero}, {Kaper}, {Langer}, {Lennon}, {Ma{\'{\i}}z
  Apell{\'a}niz}, {Markova}, {Morrell}, {Monaco}, and {Vink}]{Taylor11}
{Taylor}, W.D.; {Evans}, C.J.; {Sana}, H.; {Walborn}, N.R.; {de Mink}, S.E.;
  {Stroud}, V.E.; {Alvarez-Candal}, A.; {Barb{\'a}}, R.H.; {Bestenlehner},
  J.M.; {Bonanos}, A.Z. et al.
\newblock {The VLT-FLAMES Tarantula Survey. II. R139 revealed as a massive
  binary system}.
\newblock {\em \aap} {\bf 2011}, {\em 530},~L10,
  \href{http://xxx.lanl.gov/abs/1103.5387}{{\normalfont
  [arXiv:astro-ph.SR/1103.5387]}}.
\newblock
  doi:{\changeurlcolor{black}\href{https://doi.org/10.1051/0004-6361/201116785}{\detokenize{10.1051/0004-6361/201116785}}}.

\bibitem[{Vink} \em{et~al.}(2001){Vink}, {de Koter}, and {Lamers}]{Vink01}
{Vink}, J.S.; {de Koter}, A.; {Lamers}, H.J.G.L.M.
\newblock {Mass-loss predictions for O and B stars as a function of
  metallicity}.
\newblock {\em \aap} {\bf 2001}, {\em 369},~574--588,
  \href{http://xxx.lanl.gov/abs/astro-ph/0101509}{{\normalfont
  [astro-ph/0101509]}}.
\newblock
  doi:{\changeurlcolor{black}\href{https://doi.org/10.1051/0004-6361:20010127}{\detokenize{10.1051/0004-6361:20010127}}}.

\bibitem[{Clark} \em{et~al.}(2015){Clark}, {Bartlett}, {Broos}, {Townsley},
  {Taylor}, {Walborn}, {Bird}, {Sana}, {de Mink}, {Dufton}, {Evans}, {Langer},
  {Ma{\'{\i}}z Apell{\'a}niz}, {Schneider}, and {Soszy{\'n}ski}]{Clark15}
{Clark}, J.S.; {Bartlett}, E.S.; {Broos}, P.S.; {Townsley}, L.K.; {Taylor},
  W.D.; {Walborn}, N.R.; {Bird}, A.J.; {Sana}, H.; {de Mink}, S.E.; {Dufton},
  P.L. et al.
\newblock {The VLT-FLAMES Tarantula survey. XX. The nature of the X-ray bright
  emission-line star VFTS 399}.
\newblock {\em \aap} {\bf 2015}, {\em 579},~A131,
  \href{http://xxx.lanl.gov/abs/1503.00930}{{\normalfont
  [arXiv:astro-ph.SR/1503.00930]}}.
\newblock
  doi:{\changeurlcolor{black}\href{https://doi.org/10.1051/0004-6361/201424427}{\detokenize{10.1051/0004-6361/201424427}}}.

\bibitem[{Castor} \em{et~al.}(1975){Castor}, {Abbott}, and {Klein}]{Castor75}
{Castor}, J.I.; {Abbott}, D.C.; {Klein}, R.I.
\newblock {Radiation-driven winds in Of stars}.
\newblock {\em \apj} {\bf 1975}, {\em 195},~157--174.
\newblock
  doi:{\changeurlcolor{black}\href{https://doi.org/10.1086/153315}{\detokenize{10.1086/153315}}}.

\bibitem[{Mokiem} \em{et~al.}(2007){Mokiem}, {de Koter}, {Vink}, {Puls},
  {Evans}, {Smartt}, {Crowther}, {Herrero}, {Langer}, {Lennon}, {Najarro}, and
  {Villamariz}]{Mokiem07}
{Mokiem}, M.R.; {de Koter}, A.; {Vink}, J.S.; {Puls}, J.; {Evans}, C.J.;
  {Smartt}, S.J.; {Crowther}, P.A.; {Herrero}, A.; {Langer}, N.; {Lennon},
  D.J. et al.
\newblock {The empirical metallicity dependence of the mass-loss rate of O- and
  early B-type stars}.
\newblock {\em \aap} {\bf 2007}, {\em 473},~603--614,
  \href{http://xxx.lanl.gov/abs/0708.2042}{{\normalfont [0708.2042]}}.
\newblock
  doi:{\changeurlcolor{black}\href{https://doi.org/10.1051/0004-6361:20077545}{\detokenize{10.1051/0004-6361:20077545}}}.

\bibitem[{Prinja} \em{et~al.}(1990){Prinja}, {Barlow}, and {Howarth}]{Prinja90}
{Prinja}, R.K.; {Barlow}, M.J.; {Howarth}, I.D.
\newblock {Terminal velocities for a large sample of O stars, B supergiants,
  and Wolf-Rayet stars}.
\newblock {\em \apj} {\bf 1990}, {\em 361},~607--620.
\newblock
  doi:{\changeurlcolor{black}\href{https://doi.org/10.1086/169224}{\detokenize{10.1086/169224}}}.

\bibitem[{Roman-Duval} \em{et~al.}(2019){Roman-Duval}, {Jenkins}, {Williams},
  {Tchernyshyov}, {Gordon}, {Meixner}, {Hagen}, {Peek}, {Sandstrom}, {Werk},
  and {Yanchulova Merica-Jones}]{RomanDuval19}
{Roman-Duval}, J.; {Jenkins}, E.B.; {Williams}, B.; {Tchernyshyov}, K.;
  {Gordon}, K.; {Meixner}, M.; {Hagen}, L.; {Peek}, J.; {Sandstrom}, K.;
  {Werk}, J.; {Yanchulova Merica-Jones}, P.
\newblock {METAL: The Metal Evolution, Transport, and Abundance in the Large
  Magellanic Cloud Hubble Program. I. Overview and Initial Results}.
\newblock {\em \apj} {\bf 2019}, {\em 871},~151,
  \href{http://xxx.lanl.gov/abs/1901.06027}{{\normalfont [1901.06027]}}.
\newblock
  doi:{\changeurlcolor{black}\href{https://doi.org/10.3847/1538-4357/aaf8bb}{\detokenize{10.3847/1538-4357/aaf8bb}}}.

\bibitem[{Schmutz} \em{et~al.}(1989){Schmutz}, {Hamann}, and
  {Wessolowski}]{Schmutz89}
{Schmutz}, W.; {Hamann}, W.R.; {Wessolowski}, U.
\newblock {Spectral analysis of 30 Wolf-Rayet stars}.
\newblock {\em \aap} {\bf 1989}, {\em 210},~236--248.

\bibitem[{St.-Louis} \em{et~al.}(1988){St.-Louis}, {Moffat}, {Drissen},
  {Bastien}, and {Robert}]{StLouis88}
{St.-Louis}, N.; {Moffat}, A.F.J.; {Drissen}, L.; {Bastien}, P.; {Robert}, C.
\newblock {Polarization variability among Wolf-Rayet stars. III - A new way to
  derive mass-loss rates for Wolf-Rayet stars in binary systems}.
\newblock {\em \apj} {\bf 1988}, {\em 330},~286--304.
\newblock
  doi:{\changeurlcolor{black}\href{https://doi.org/10.1086/166472}{\detokenize{10.1086/166472}}}.

\bibitem[{Hillier}(1991)]{Hillier91}
{Hillier}, D.J.
\newblock {The effects of electron scattering and wind clumping for early
  emission line stars}.
\newblock {\em \aap} {\bf 1991}, {\em 247},~455--468.

\bibitem[{Fullerton} \em{et~al.}(2006){Fullerton}, {Massa}, and
  {Prinja}]{Fullerton06}
{Fullerton}, A.W.; {Massa}, D.L.; {Prinja}, R.K.
\newblock {The Discordance of Mass-Loss Estimates for Galactic O-Type Stars}.
\newblock {\em \apj} {\bf 2006}, {\em 637},~1025--1039,
  \href{http://xxx.lanl.gov/abs/astro-ph/0510252}{{\normalfont
  [astro-ph/0510252]}}.
\newblock
  doi:{\changeurlcolor{black}\href{https://doi.org/10.1086/498560}{\detokenize{10.1086/498560}}}.

\bibitem[{Sundqvist} and {Puls}(2018)]{Sundqvist18}
{Sundqvist}, J.O.; {Puls}, J.
\newblock {Atmospheric NLTE models for the spectroscopic analysis of blue stars
  with winds. IV. Porosity in physical and velocity space}.
\newblock {\em \aap} {\bf 2018}, {\em 619},~A59,
  \href{http://xxx.lanl.gov/abs/1805.11010}{{\normalfont
  [arXiv:astro-ph.SR/1805.11010]}}.
\newblock
  doi:{\changeurlcolor{black}\href{https://doi.org/10.1051/0004-6361/201832993}{\detokenize{10.1051/0004-6361/201832993}}}.

\bibitem[{Lamers} \em{et~al.}(1995){Lamers}, {Snow}, and {Lindholm}]{Lamers95}
{Lamers}, H.J.G.L.M.; {Snow}, T.P.; {Lindholm}, D.M.
\newblock {Terminal Velocities and the Bistability of Stellar Winds}.
\newblock {\em \apj} {\bf 1995}, {\em 455},~269.
\newblock
  doi:{\changeurlcolor{black}\href{https://doi.org/10.1086/176575}{\detokenize{10.1086/176575}}}.

\bibitem[{Kudritzki} and {Puls}(2000)]{KudritzkiPuls00}
{Kudritzki}, R.P.; {Puls}, J.
\newblock {Winds from Hot Stars}.
\newblock {\em \araa} {\bf 2000}, {\em 38},~613--666.
\newblock
  doi:{\changeurlcolor{black}\href{https://doi.org/10.1146/annurev.astro.38.1.613}{\detokenize{10.1146/annurev.astro.38.1.613}}}.

\bibitem[{Gr{\"a}fener} \em{et~al.}(2011){Gr{\"a}fener}, {Vink}, {de Koter},
  and {Langer}]{Grafener11}
{Gr{\"a}fener}, G.; {Vink}, J.S.; {de Koter}, A.; {Langer}, N.
\newblock {The Eddington factor as the key to understand the winds of the most
  massive stars. Evidence for a {$\Gamma$}-dependence of Wolf-Rayet type mass
  loss}.
\newblock {\em \aap} {\bf 2011}, {\em 535},~A56,
  \href{http://xxx.lanl.gov/abs/1106.5361}{{\normalfont
  [arXiv:astro-ph.SR/1106.5361]}}.
\newblock
  doi:{\changeurlcolor{black}\href{https://doi.org/10.1051/0004-6361/201116701}{\detokenize{10.1051/0004-6361/201116701}}}.

\bibitem[{Bestenlehner}("in prep."){Bestenlehner}]{Bestenlehner19a}
{Bestenlehner}, J.M.
\newblock {Mass loss and the Eddington factor: an updated stellar wind theory for hot massive stars}.
\newblock {\em \mnras} {\bf in prep.}.


\bibitem[{Puls} \em{et~al.}(2008){Puls}, {Vink}, {Najarro}]{Puls08}
{Puls}, J..; {Vink}, J.S.; {Najarro}, F.
\newblock {Mass loss from hot massive stars}.
\newblock {\em \aapr} {\bf 2008}, {\em 16},~209-325,
  \href{http://xxx.lanl.gov/abs/0811.0487}{{\normalfont
  [arXiv:astro-ph.SR/0811.0487]}}.


\bibitem[{Langer}(2012)]{Langer12}
{Langer}, N.
\newblock {Presupernova Evolution of Massive Single and Binary Stars}.
\newblock {\em \araa} {\bf 2012}, {\em 50},~107--164,
  \href{http://xxx.lanl.gov/abs/1206.5443}{{\normalfont
  [arXiv:astro-ph.SR/1206.5443]}}.
\newblock
  doi:{\changeurlcolor{black}\href{https://doi.org/10.1146/annurev-astro-081811-125534}{\detokenize{10.1146/annurev-astro-081811-125534}}}.




\bibitem[{Yusof} \em{et~al.}(2013){Yusof}, {Hirschi}, {Meynet}, {Crowther},
  {Ekstr{\"o}m}, {Frischknecht}, {Georgy}, {Abu Kassim}, and
  {Schnurr}]{Yusof13}
{Yusof}, N.; {Hirschi}, R.; {Meynet}, G.; {Crowther}, P.A.; {Ekstr{\"o}m}, S.;
  {Frischknecht}, U.; {Georgy}, C.; {Abu Kassim}, H.; {Schnurr}, O.
\newblock {Evolution and fate of very massive stars}.
\newblock {\em \mnras} {\bf 2013}, {\em 433},~1114--1132,
  \href{http://xxx.lanl.gov/abs/1305.2099}{{\normalfont
  [arXiv:astro-ph.SR/1305.2099]}}.
\newblock
  doi:{\changeurlcolor{black}\href{https://doi.org/10.1093/mnras/stt794}{\detokenize{10.1093/mnras/stt794}}}.

\bibitem[{Johnson} \em{et~al.}(2017){Johnson}, {Rigby}, {Sharon}, {Gladders},
  {Florian}, {Bayliss}, {Wuyts}, {Whitaker}, {Livermore}, and
  {Murray}]{Johnson17}
{Johnson}, T.L.; {Rigby}, J.R.; {Sharon}, K.; {Gladders}, M.D.; {Florian}, M.;
  {Bayliss}, M.B.; {Wuyts}, E.; {Whitaker}, K.E.; {Livermore}, R.; {Murray},
  K.T.
\newblock {Star Formation at z = 2.481 in the Lensed Galaxy SDSS J1110+6459:
  Star Formation Down to 30 pc Scales}.
\newblock {\em \apjl} {\bf 2017}, {\em 843},~L21,
  \href{http://xxx.lanl.gov/abs/1707.00706}{{\normalfont [1707.00706]}}.
\newblock
  doi:{\changeurlcolor{black}\href{https://doi.org/10.3847/2041-8213/aa7516}{\detokenize{10.3847/2041-8213/aa7516}}}.

\bibitem[{Micheva} \em{et~al.}(2017){Micheva}, {Oey}, {Jaskot}, and
  {James}]{Micheva17}
{Micheva}, G.; {Oey}, M.S.; {Jaskot}, A.E.; {James}, B.L.
\newblock {Mrk 71/NGC 2366: The Nearest Green Pea Analog}.
\newblock {\em \apj} {\bf 2017}, {\em 845},~165,
  \href{http://xxx.lanl.gov/abs/1704.01678}{{\normalfont [1704.01678]}}.
\newblock
  doi:{\changeurlcolor{black}\href{https://doi.org/10.3847/1538-4357/aa830b}{\detokenize{10.3847/1538-4357/aa830b}}}.

\bibitem[{Pellegrini} \em{et~al.}(2010){Pellegrini}, {Baldwin}, and
  {Ferland}]{Pellegrini10}
{Pellegrini}, E.W.; {Baldwin}, J.A.; {Ferland}, G.J.
\newblock {Structure and Feedback in 30 Doradus. I. Observations}.
\newblock {\em \apjs} {\bf 2010}, {\em 191},~160--178,
  \href{http://xxx.lanl.gov/abs/1009.4948}{{\normalfont [1009.4948]}}.
\newblock
  doi:{\changeurlcolor{black}\href{https://doi.org/10.1088/0067-0049/191/1/160}{\detokenize{10.1088/0067-0049/191/1/160}}}.

\bibitem[{Vacca} \em{et~al.}(1995){Vacca}, {Robert}, {Leitherer}, and
  {Conti}]{Vacca95}
{Vacca}, W.D.; {Robert}, C.; {Leitherer}, C.; {Conti}, P.S.
\newblock {The stellar content of 30 doradus derived from spatially integrated
  ultraviolet spectra: A test of spectral synthesis models}.
\newblock {\em \apj} {\bf 1995}, {\em 444},~647--662.
\newblock
  doi:{\changeurlcolor{black}\href{https://doi.org/10.1086/175637}{\detokenize{10.1086/175637}}}.

\bibitem[{Smith} \em{et~al.}(2016){Smith}, {Crowther}, {Calzetti}, and
  {Sidoli}]{Smith16}
{Smith}, L.J.; {Crowther}, P.A.; {Calzetti}, D.; {Sidoli}, F.
\newblock {The Very Massive Star Content of the Nuclear Star Clusters in NGC
  5253}.
\newblock {\em \apj} {\bf 2016}, {\em 823},~38,
  \href{http://xxx.lanl.gov/abs/1603.06974}{{\normalfont [1603.06974]}}.
\newblock
  doi:{\changeurlcolor{black}\href{https://doi.org/10.3847/0004-637X/823/1/38}{\detokenize{10.3847/0004-637X/823/1/38}}}.

\bibitem[{Levesque} \em{et~al.}(2012){Levesque}, {Leitherer}, {Ekstrom},
  {Meynet}, and {Schaerer}]{Levesque12}
{Levesque}, E.M.; {Leitherer}, C.; {Ekstrom}, S.; {Meynet}, G.; {Schaerer}, D.
\newblock {The Effects of Stellar Rotation. I. Impact on the Ionizing Spectra
  and Integrated Properties of Stellar Populations}.
\newblock {\em \apj} {\bf 2012}, {\em 751},~67,
  \href{http://xxx.lanl.gov/abs/1203.5109}{{\normalfont
  [arXiv:astro-ph.SR/1203.5109]}}.
\newblock
  doi:{\changeurlcolor{black}\href{https://doi.org/10.1088/0004-637X/751/1/67}{\detokenize{10.1088/0004-637X/751/1/67}}}.

\bibitem[{Eldridge} \em{et~al.}(2017){Eldridge}, {Stanway}, {Xiao},
  {McClelland}, {Taylor}, {Ng}, {Greis}, and {Bray}]{Eldridge17}
{Eldridge}, J.J.; {Stanway}, E.R.; {Xiao}, L.; {McClelland}, L.A.S.; {Taylor},
  G.; {Ng}, M.; {Greis}, S.M.L.; {Bray}, J.C.
\newblock {Binary Population and Spectral Synthesis Version 2.1: Construction,
  Observational Verification, and New Results}.
\newblock {\em \pasa} {\bf 2017}, {\em 34},~e058,
  \href{http://xxx.lanl.gov/abs/1710.02154}{{\normalfont
  [arXiv:astro-ph.SR/1710.02154]}}.
\newblock
  doi:{\changeurlcolor{black}\href{https://doi.org/10.1017/pasa.2017.51}{\detokenize{10.1017/pasa.2017.51}}}.


\bibitem[{Moffat} \em{et~al.}(2002){Moffat}, {Corcoran}, {Stevens},
  {Skalkowski}, {Marchenko}, {M{\"u}cke}, {Ptak}, {Koribalski}, {Brenneman},
  {Mushotzky}, {Pittard}, {Pollock}, and {Brandner}]{Moffat02}
{Moffat}, A.F.J.; {Corcoran}, M.F.; {Stevens}, I.R.; {Skalkowski}, G.;
  {Marchenko}, S.V.; {M{\"u}cke}, A.; {Ptak}, A.; {Koribalski}, B.S.;
  {Brenneman}, L.; {Mushotzky}, R. et al.
\newblock {Galactic Starburst NGC 3603 from X-Rays to Radio}.
\newblock {\em \apj} {\bf 2002}, {\em 573},~191--198.
\newblock
  doi:{\changeurlcolor{black}\href{https://doi.org/10.1086/340491}{\detokenize{10.1086/340491}}}.



\bibitem[{Melena} \em{et~al.}(2008){Melena}, {Massey}, {Morrell}, and
  {Zangari}]{Melena08}
{Melena}, N.W.; {Massey}, P.; {Morrell}, N.I.; {Zangari}, A.M.
\newblock {The Massive Star Content of NGC 3603}.
\newblock {\em \aj} {\bf 2008}, {\em 135},~878--891,
  \href{http://xxx.lanl.gov/abs/0712.2621}{{\normalfont [0712.2621]}}.
\newblock
  doi:{\changeurlcolor{black}\href{https://doi.org/10.1088/0004-6256/135/3/878}{\detokenize{10.1088/0004-6256/135/3/878}}}.


\bibitem[{Ramachandran} \em{et~al.}(2019){Ramachandran}, {Hamann}, {Oskinova},
  {Gallagher}, {Hainich}, {Shenar}, {Sander}, {Todt}, and
  {Fulmer}]{Ramachandran19}
{Ramachandran}, V.; {Hamann}, W.R.; {Oskinova}, L.M.; {Gallagher}, J.S.;
  {Hainich}, R.; {Shenar}, T.; {Sander}, A.A.C.; {Todt}, H.; {Fulmer}, L.
\newblock {Testing massive star evolution, star formation history, and feedback
  at low metallicity. Spectroscopic analysis of OB stars in the SMC Wing}.
\newblock {\em \aap} {\bf 2019}, {\em 625},~A104,
  \href{http://xxx.lanl.gov/abs/1903.01762}{{\normalfont
  [arXiv:astro-ph.SR/1903.01762]}}.
\newblock
  doi:{\changeurlcolor{black}\href{https://doi.org/10.1051/0004-6361/201935365}{\detokenize{10.1051/0004-6361/201935365}}}.

\bibitem[{Baldwin} \em{et~al.}(1981){Baldwin}, {Phillips}, and
  {Terlevich}]{Baldwin81}
{Baldwin}, J.A.; {Phillips}, M.M.; {Terlevich}, R.
\newblock {Classification parameters for the emission-line spectra of
  extragalactic objects}.
\newblock {\em \pasp} {\bf 1981}, {\em 93},~5--19.
\newblock
  doi:{\changeurlcolor{black}\href{https://doi.org/10.1086/130766}{\detokenize{10.1086/130766}}}.

  
\bibitem[{Steidel} \em{et~al.}(2014){Steidel}, {Rudie}, {Strom}, {Pettini},
  {Reddy}, {Shapley}, {Trainor}, {Erb}, {Turner}, {Konidaris}, {Kulas}, {Mace},
  {Matthews}, and {McLean}]{Steidel14}
{Steidel}, C.C.; {Rudie}, G.C.; {Strom}, A.L.; {Pettini}, M.; {Reddy}, N.A.;
  {Shapley}, A.E.; {Trainor}, R.F.; {Erb}, D.K.; {Turner}, M.L.; {Konidaris},
  N.P. et al.
\newblock {Strong Nebular Line Ratios in the Spectra of z \~{} 2-3 Star Forming
  Galaxies: First Results from KBSS-MOSFIRE}.
\newblock {\em \apj} {\bf 2014}, {\em 795},~165,
  \href{http://xxx.lanl.gov/abs/1405.5473}{{\normalfont [1405.5473]}}.
\newblock
  doi:{\changeurlcolor{black}\href{https://doi.org/10.1088/0004-637X/795/2/165}{\detokenize{10.1088/0004-637X/795/2/165}}}.

\bibitem[{Izotov} \em{et~al.}(2016{\natexlab{a}}){Izotov}, {Orlitov{\'a}},
  {Schaerer}, {Thuan}, {Verhamme}, {Guseva}, and {Worseck}]{Izotov16a}
{Izotov}, Y.I.; {Orlitov{\'a}}, I.; {Schaerer}, D.; {Thuan}, T.X.; {Verhamme},
  A.; {Guseva}, N.G.; {Worseck}, G.
\newblock {Eight per cent leakage of Lyman continuum photons from a compact,
  star-forming dwarf galaxy}.
\newblock {\em \nat} {\bf 2016}, {\em 529},~178--180,
  \href{http://xxx.lanl.gov/abs/1601.03068}{{\normalfont [1601.03068]}}.
\newblock
  doi:{\changeurlcolor{black}\href{https://doi.org/10.1038/nature16456}{\detokenize{10.1038/nature16456}}}.

\bibitem[{Izotov} \em{et~al.}(2016{\natexlab{b}}){Izotov}, {Schaerer}, {Thuan},
  {Worseck}, {Guseva}, {Orlitov{\'a}}, and {Verhamme}]{Izotov16b}
{Izotov}, Y.I.; {Schaerer}, D.; {Thuan}, T.X.; {Worseck}, G.; {Guseva}, N.G.;
  {Orlitov{\'a}}, I.; {Verhamme}, A.
\newblock {Detection of high Lyman continuum leakage from four low-redshift
  compact star-forming galaxies}.
\newblock {\em \mnras} {\bf 2016}, {\em 461},~3683--3701,
  \href{http://xxx.lanl.gov/abs/1605.05160}{{\normalfont [1605.05160]}}.
\newblock
  doi:{\changeurlcolor{black}\href{https://doi.org/10.1093/mnras/stw1205}{\detokenize{10.1093/mnras/stw1205}}}.

\bibitem[{Steidel} \em{et~al.}(2016){Steidel}, {Strom}, {Pettini}, {Rudie},
  {Reddy}, and {Trainor}]{Steidel16}
{Steidel}, C.C.; {Strom}, A.L.; {Pettini}, M.; {Rudie}, G.C.; {Reddy}, N.A.;
  {Trainor}, R.F.
\newblock {Reconciling the Stellar and Nebular Spectra of High-redshift
  Galaxies}.
\newblock {\em \apj} {\bf 2016}, {\em 826},~159,
  \href{http://xxx.lanl.gov/abs/1605.07186}{{\normalfont [1605.07186]}}.
\newblock
  doi:{\changeurlcolor{black}\href{https://doi.org/10.3847/0004-637X/826/2/159}{\detokenize{10.3847/0004-637X/826/2/159}}}.


\bibitem[{Kennicutt} \em{et~al.}(2008){Kennicutt}, {Lee}, {Funes}, {J.},
  {Sakai}, and {Akiyama}]{Kennicutt08}
{Kennicutt}, Jr., R.C.; {Lee}, J.C.; {Funes}, J.G.; {J.}, S.; {Sakai}, S.;
  {Akiyama}, S.
\newblock {An H{$\alpha$} Imaging Survey of Galaxies in the Local 11 Mpc
  Volume}.
\newblock {\em \apjs} {\bf 2008}, {\em 178},~247--279,
  \href{http://xxx.lanl.gov/abs/0807.2035}{{\normalfont [0807.2035]}}.
\newblock
  doi:{\changeurlcolor{black}\href{https://doi.org/10.1086/590058}{\detokenize{10.1086/590058}}}.

\bibitem[{van Zee} \em{et~al.}(2006){van Zee}, {Skillman}, and
  {Haynes}]{vanZee06}
{van Zee}, L.; {Skillman}, E.D.; {Haynes}, M.P.
\newblock {Oxygen and Nitrogen in Leo A and GR 8}.
\newblock {\em \apj} {\bf 2006}, {\em 637},~269--282,
  \href{http://xxx.lanl.gov/abs/astro-ph/0509678}{{\normalfont
  [arXiv:astro-ph/astro-ph/0509678]}}.
\newblock
  doi:{\changeurlcolor{black}\href{https://doi.org/10.1086/498298}{\detokenize{10.1086/498298}}}.

\bibitem[{Saviane} \em{et~al.}(2002){Saviane}, {Rizzi}, {Held}, {Bresolin}, and
  {Momany}]{Saviane02}
{Saviane}, I.; {Rizzi}, L.; {Held}, E.V.; {Bresolin}, F.; {Momany}, Y.
\newblock {New abundance measurements in UKS 1927-177, a very metal-poor galaxy
  in the Local Group}.
\newblock {\em \aap} {\bf 2002}, {\em 390},~59--64,
  \href{http://xxx.lanl.gov/abs/astro-ph/0205355}{{\normalfont
  [arXiv:astro-ph/astro-ph/0205355]}}.
\newblock
  doi:{\changeurlcolor{black}\href{https://doi.org/10.1051/0004-6361:20020750}{\detokenize{10.1051/0004-6361:20020750}}}.

\bibitem[{Garcia} \em{et~al.}(2019){Garcia}, {Herrero}, {Najarro}, {Camacho},
  and {Lorenzo}]{Garcia19}
{Garcia}, M.; {Herrero}, A.; {Najarro}, F.; {Camacho}, I.; {Lorenzo}, M.
\newblock {Ongoing star formation at the outskirts of Sextans A: spectroscopic
  detection of early O-type stars}.
\newblock {\em \mnras} {\bf 2019}, {\em 484},~422--430,
  \href{http://xxx.lanl.gov/abs/1901.02466}{{\normalfont [1901.02466]}}.
\newblock
  doi:{\changeurlcolor{black}\href{https://doi.org/10.1093/mnras/sty3503}{\detokenize{10.1093/mnras/sty3503}}}.

\bibitem[{Walborn} \em{et~al.}(1985){Walborn}, {Nichols-Bohlin}, and
  {Panek}]{Walborn85}
{Walborn}, N.R.; {Nichols-Bohlin}, J.; {Panek}, R.J.
\newblock {International Ultraviolet Explorer Atlas of O-type Spectra from 1200
  to 1900 {\AA}.}
\newblock {\em NASA Reference Publication} {\bf 1985}, {\em 1155}.

\end{thebibliography}



\end{document}